# The MYStIX InfraRed-Excess Source Catalog

Matthew S. Povich,[1,2] Michael A. Kuhn,[2] Konstantin V. Getman,[2] Heather A. Busk,[2] Eric D. Feigelson,[2] Patrick S. Broos,[2] Leisa K. Townsley,[2] Robert R. King,[3] & Tim Naylor[3]

## ABSTRACT

The MYStIX project (**M**assive **Y**oung Star-Forming Complex **St**udy in **I**nfrared and **X**-rays) provides a comparative study of 20 Galactic massive star-forming complexes ($d = 0.4$ to 3.6 kpc). Probable stellar members in each target complex are identified using X-ray and/or infrared data via two pathways: (1) X-ray detections of young/massive stars with coronal activity/strong winds; or (2) infrared excess (IRE) selection of young stellar objects (YSOs) with circumstellar disks and/or protostellar envelopes. We present the methodology for the second pathway, using *Spitzer*/IRAC, 2MASS, and UKIRT imaging and photometry. Although IRE selection of YSOs is well-trodden territory, MYStIX presents unique challenges. The target complexes range from relatively nearby clouds in uncrowded fields located toward the outer Galaxy (e.g. NGC 2264, the Flame Nebula) to more distant, massive complexes situated along complicated, inner Galaxy sightlines (e.g. NGC 6357, M17). We combine IR spectral energy distribution (SED) fitting with IR color cuts and spatial clustering analysis to identify IRE sources and isolate probable YSO members in each MYStIX target field from the myriad types of contaminating sources that can resemble YSOs: extragalactic sources, evolved stars, nebular knots, and even unassociated foreground/background YSOs. Applying our methodology consistently across 18 of the target complexes, we produce the MYStIX IRE Source (MIRES) Catalog comprising 20,719 sources, including 8686 probable stellar members of the MYStIX target complexes. We also classify the SEDs of 9365 IR counterparts to MYStIX X-ray sources to assist the first pathway, the identification of X-ray–detected stellar members. The MIRES catalog provides a foundation for follow-up studies of diverse phenomena related to massive star cluster formation, including protostellar outflows, circumstellar disks, and sequential star formation triggered by massive star feedback processes.

---

[1] California State Polytechnic University, 3801 West Temple Ave, Pomona, CA 91768; mspovich@csupomona.edu

[2] Department of Astronomy and Astrophysics, The Pennsylvania State University, 525 Davey Lab, University Park, PA 16802

[3] School of Physics, University of Exeter, Exeter EX4 4QL, UK

*Subject headings:* infrared: stars — Methods: data analysis — open clusters and associations — planetary systems: protoplanetary disks — Stars: pre-main-sequence — Stars: protostars

## 1. Introduction

The **M**assive **Y**oung star-forming complex **St**udy in **I**nfrared and **X**-rays (MYStIX) project, described by Feigelson et al. (2013), provides a comprehensive, parallel study of 20 Galactic massive star-forming regions (MSFRs; $d = 0.4$ to 3.6 kpc). The core data products of MYStIX are tables of "MYStIX Probable Complex Members" (MPCMs) in each target MSFR, compiled by Broos et al. (2013). MPCMs are identified using a combination of X-ray imaging data from the *Chandra X-ray Observatory* and infrared (IR) data from the United Kingdom Infrared Telescope (UKIRT), the Two-Micron All-Sky Survey (2MASS), and the *Spitzer Space Telescope*. Young, pre-main–sequence (pre-MS) stars with convection-driven coronal flaring activity (Feigelson et al. 2002; Flaccomio et al. 2003; Preibisch et al. 2005; Güdel et al. 2007) and massive, OB stars with strong stellar winds (e.g. Harnden et al. 1979; Gagné et al. 2011) produce bright X-ray emission that allows them to be isolated in high-resolution X-ray images from the potentially overwhelming field-star and nebular contamination that plague optical/IR images of young, massive Galactic star-forming regions. X-ray observations are thus efficient probes of MPCMs both with and without circumstellar disks, penetrating obscuring absorption columns equivalent to tens of magnitudes of optical extinction $A_V$.

There are, however, some important limitations to basing MPCM identification on X-ray selection alone. X-ray emission from pre-MS stars is generally variable, and there is a wide scatter in the stellar $L_X/L_{\rm bol}$ correlation (Preibisch et al. 2005; Telleschi et al. 2007; Güdel & Nazé 2009). Hence stars that happen to be intrinsically less luminous or in a low state may be missed by the relatively shallow X-ray integrations available to MYStIX. In addition, both classical T Tauri stars with disks and protostars still accreting from infalling envelopes (to which we will refer collectively as young stellar objects, or YSOs) are observed to be somewhat less luminous in X-rays compared to diskless, weak-lined T Tauri stars (Telleschi et al. 2007; Prisinzano et al. 2008), and are usually subject to greater soft X-ray absorption. YSOs are thus underrepresented in X-ray surveys compared to older pre-MS stars. The dusty disks and/or envelopes surrounding YSOs reprocess stellar radiation, producing IR excess (IRE) emission, hence *Spitzer* observations in particular are highly sensitive to precisely the stellar populations that *Chandra* may miss (Povich et al. 2011). The complementarity between X-ray and IRE detection is a crucial motivation in our search for a more comprehensive survey of young stellar populations in MYStIX. To that end, we present the **M**YStIX **I**nfra**R**ed **E**xcess **S**ource catalog (MIRES), a compilation of IRE sources identified in wide-field IR survey images of the MYStIX target MSFRs.

Since its launch in 2003 (Werner et al. 2004), *Spitzer* has been an engine for YSO detection and characterization (see Allen et al. 2004; Robitaille et al. 2008; Gutermuth et al. 2009; Povich et al. 2009, 2011, and many others). Over this past decade, there have been many variations on the basic methodology used for IRE identification, based on broadband colors, spectral indices, spectral energy distribution (SED) fitting, or some combination. Against this backdrop of recent history, MIRES presents a novel, unique challenge. MYStIX requires a single methodology to catalog IRE sources and establish their probable membership in target regions that run the gamut from relatively nearby clouds in uncrowded fields presenting sightlines toward the outer Galaxy (e.g. NGC 2264 and the Flame Nebula) to more distant MSFRs situated along complicated, inner Galaxy sightlines (e.g. NGC 6334, NGC 6357, and the Trifid Nebula). In deep *Spitzer* observations toward the outer Galaxy or away from the Galactic midplane, the principal sources of contamination to YSO searches are intrinsically red, unresolved extragalactic sources, namely starburst galaxies and obscured active galactic nuclei (Gutermuth et al. 2009, hereafter G09). By contrast, the inner Galaxy MYStIX targets are observed against large field populations of highly-reddened giants, dust-rich asymptotic giant branch stars, and even YSOs from multiple star-forming clouds overlapping along the complicated sightlines through the Galactic disk, all of which conspire to produce significant contamination (Robitaille et al. 2008; Povich et al. 2009). In constructing MIRES, we have combined best practices from the literature to optimize identification of IRE sources while separating probable YSO MPCMs from various types of contaminants. Our approach is conservative, opting to exclude likely YSOs if their properties resemble those of contaminant populations. Nonetheless, we produce one of the largest reliable catalogs of YSOs associated with Galactic star-forming regions compiled to date.

This contribution is intended to serve primarily as a description of MIRES as both a catalog (published as an accompanying online table) and a methodology for identifying YSOs from broadband photometric data. For the basic science goals and target selection of MYStIX, we refer the reader to Feigelson et al. (2013). MIRES includes 18 of the 20 MYStIX targets, as the remaining two, the Carina Nebula Complex and the Orion Molecular Cloud Complex, each have large, high-reliability YSO catalogs recently published (Povich et al. 2011; Megeath et al. 2012). New scientific results based on MYStIX and MIRES will appear in future papers. The remainder of this paper is organized as follows: In Section 2 we describe the various IR photometry catalogs used for MIRES, and in Section 3 we present the detailed analysis methodology used to select IRE sources for MIRES. The MIRES catalog itself (provided as an accompanying online table) is described in Section 4. In Section 5 we detail our strategy for classifying MIRES as probable stellar members of their parent MSFRs. High-level results for MIRES as a whole are summarized in Section 6. We also include two appendices: in Appendix B we describe qualitative results for each of the 18 individual MSFRs, and in Appendix A we use the MIRES methodology to classify IR counterparts to MYStIX X-ray sources (including an accompanying online table).

## 2. Infrared Source Catalogs

The basic input data for MIRES were near-IR (NIR) and mid-IR (MIR) photometric catalogs. We also use NIR and MIR images and mosaics for visualizing the point source populations with respect to various nebular structures. We provide high-level descriptions of each input catalog below. For more detailed information, we refer the reader to the primary sources cited for each catalog.

### 2.1. *Spitzer*/IRAC

Our selection criteria for circumstellar material rely on IRE emission detected in two or more of the four MIR bands available to the Infrared Array Camera (IRAC; Fazio et al. 2004) on the *Spitzer Space Telescope*. The IRAC bands (cryogenic mission phase) are centered at 3.6, 4.5, 5.8, and 8.0 $\mu$m, and we will henceforth use the notation [3.6], [4.5], [5.8], and [8.0], respectively when referring to a specific IRAC band or photometric magnitudes measured from it.

The Galactic Legacy Infrared Mid-Plane Survey Extraordinaire (GLIMPSE; Benjamin et al. 2003), and 3 follow-up survey programs using the GLIMPSE observing strategies and data analysis pipelines (GLIMPSE II, GLIMPSE 3D, and the Vela–Carina Survey) observed 8 of the 18 MYStIX target MSFRs (see Table 1). The GLIMPSE photometry pipeline provides a highly-reliable point source "Catalog" that is a subset of a more complete point source "Archive". For IRE source selection, we use the highly-reliable Catalog exclusively. For analysis of MIR counterparts matched to MYStIX X-ray sources (Naylor et al. 2013) we use the more-complete Archive (see Appendix A for details).

The wide-area sky coverage provided by the GLIMPSE surveys allowed us to define large search fields for MIRES, approaching the full extent of 8.0 $\mu$m nebulosity (a qualitative tracer of molecular clouds) associated with these MSFRs. These wider MIRES fields are generally much larger than the MYStIX X-ray fields, which allows us (1) to identify centers of star-forming activity that did not happen to fall within the X-ray observations and (2) to define off-target, "control" regions and establish a baseline density for contaminating field sources that masquerade as MSFR members with IRE (see Section 5.1).

Kuhn et al. (2013b, hereafter K13) performed MIR point-source photometry on archival IRAC data for the 10 other MYStIX targets used in MIRES that were not covered by one of the GLIMPSE surveys (see Table 1). K13 modeled their catalog structure on the GLIMPSE pipeline and similarly produced both highly-reliable Catalog (their primary published data product) and more-complete Archive source lists. There are, however, important differences from GLIMPSE in both the data analyzed by K13 and the photometry pipeline itself:

1. *Photometric depth.* In general, the archival data analyzed by K13 were deeper integrations than the GLIMPSE surveys. Faint, extragalactic sources are rare in the GLIMPSE Catalogs but prevalent in the K13 Catalogs.

2. *Fields-of-view.* The archival IRAC data came from various programs, and the size of the fields-of-view differ greatly among the targets.

3. *Source detection and extraction.* K13 used a point-source detection algorithm with a less stringent roundness criterion compared to the GLIMPSE pipeline. K13 performed aperture photometry on mosaic images combining all available IRAC data, while the GLIMPSE pipeline performed point-spread-function (PSF) fitting photometry on the individual IRAC frames. K13 note that while their pipeline tends to detect point sources that the GLIMPSE pipeline would miss, it is more susceptible to false-positives, especially in the [5.8] and [8.0] bands where the background nebulosity is brightest.

For two targets, K13 compare the results of their photometry pipeline to those of the GLIMPSE pipeline. A custom run of the GLIMPSE pipeline was performed on archival data for W3 (M. R. Meade and B. L. Babler, private communication). K13 produced sourcelists from the IRAC high-dynamic-range GTO observation of the central regions of M17, which was also included (with wider coverage) in the GLIMPSE survey. Both of these targets include luminous H II regions that produce very bright, complex nebular emission on multiple spatial scales. K13 found that in these cases their pipeline produced a significant number of spurious detections (point 3 above) at [5.8] and [8.0]. As these spurious detections produced an unacceptably high fraction of false IRE sources in W3 and M17, for these targets we use the GLIMPSE pipeline Catalogs. We note that the GLIMPSE pipeline was optimized for the (conservative) detection of crowded sources against complicated nebular background emission, and, for the purposes of MIRES, reliability (in the sense of minimizing false-positives) takes priority over completeness (minimizing false negatives).

## 2.2. 2MASS

*2MASS* (Skrutskie et al. 2006) imaging is well-matched to the 2 resolution of *Spitzer*/IRAC and provides an all-sky, broadband $JHK_S$ photometric catalog covering the NIR bands immediately blueward of IRAC [3.6]. Both the GLIMPSE and K13 IRAC photometry pipelines produce 7-band catalogs with *2MASS* sources spatially matched to IRAC sources.

## 2.3. The United Kingdom Infrared Telescope (UKIRT)

Although the sensitivity of *2MASS* is nominally well-matched to that of the GLIMPSE surveys ($K_S$ and [3.6] 15.5 mag), differential extinction (the combination of interstellar and circumstellar reddening) renders many IRE sources significantly fainter at NIR wavelengths. For MYStIX targets with deeper IRAC data, *2MASS* is clearly not deep enough. We therefore incorporate $JHK_S$ photometry catalogs produced by King et al. (2013) from a combination of UKIRT Infrared Deep

Sky Survey (UKIDSS) imaging and similar observations targeting select MSFRs, where available (see King et al. (2013) and Table 1). UKIRT observations provide sub-arcsec resolution and reach $K_S$ 19 mag. For the Lagoon Nebula, NGC 6334, and NGC 6357 the MIRES search field was limited by the area of the corresponding UKIRT catalog.

### 2.4. Cross-Catalog Source Matching

As a preliminary step to our MIRES selection and analysis procedure, for each target MSFR with available UKIRT photometry, we cross-matched the relevant King et al. (2013) source catalog with the appropriate subset of the GLIMPSE or K13 highly-reliable MIR Catalog (Table 1). Because the MIR Catalogs already incorporated *2MASS* photometry, for a total of 7 bands, the goal was to replace *2MASS* with high-quality UKIRT photometry and populate missing NIR photometry with UKIRT magnitudes wherever possible. The matching was based on astrometric proximity, following the techniques of Broos et al. (2011). The steps in our matching procedure can be summarized as follows:

1. Define the common field-of-view (FOV) of the UKIRT and MIR coverage, and crop both catalogs to this common FOV, which we hereafter call "the MIRES full field."

2. Remove artifacts from the UKIRT catalogs that do not represent astrophysical objects. These include all sources with the following flags (King et al. 2013): W (calibration extractions in any band), E (near edge in $K_S$ band), and M (negative flux in $K_S$ band).

3. Perform a S/N cut in $K_S$, keeping only sources for which the photometric uncertainty $\delta K_S <$ 0.1 mag.

4. Register the MIR Catalog to bright ($K_S <$ 14 mag) sources in the UKIRT catalog, excluding saturated sources (S flag).

5. Match UKIRT sources to MIR Catalog sources using a 1 matching radius. If multiple UKIRT sources fall within the matching radius of a MIR source, then the closest is adopted as the match and the number of secondary matches is recorded.

The results of the matching procedure were evaluated using visual review of the UKIRT and MIR catalog sources on the relevant $K_S$ and 3.6 $\mu$m images, and by plotting *2MASS* $K_S$ versus UKIRT $K_S$ matched to the same MIR source. In the latter case, we found that the sources correlated tightly with the 1:1 line, with the exception of saturated UKIRT sources that were unsaturated in *2MASS* and UKIRT sources with secondary matches present, both of which skew toward larger (fainter) values of $K_S$ in UKIRT compared to *2MASS*. Generally, secondary UKIRT matches to a MIR source represent close pairs (or triplets) of NIR sources that were unresolvable at the 2 resolution of either *2MASS* or *Spitzer*/IRAC.

For the nearest MYStIX MSFR with available UKIRT data, NGC 2264, 88% of 22,363 MIR (K13) sources had UKIRT matches, only 0.4% of which were accompanied by secondary matches. This represents a best-case scenario for cross-matching. The worst-case scenarios were more distant MSFRs in the inner Galaxy, with large ( 1°) FOVs densely populated by field stars, for example NGC 6357, NGC 6334, and the Trifid Nebula. For each of these targets the above matching procedure found UKIRT matches to 95% of MIR (GLIMPSE II) sources, 15% of which were accompanied by secondary matches.

## 3. MIRES Catalog Construction

In this section we describe our methodology for identifying IRE sources among the $1.6 \times 10^6$ sources (Table 2) in our merged IRAC, *2MASS*, and UKIRT catalogs. The general strategy is best described as a series of filters to cull out various populations of contaminating sources that dominate the IR catalogs, including normally-reddened field stars, "marginal" IRE sources that depart from normal photospheric emission only at the longest wavelengths, and "bad data" sources with photometry that is inconsistent with any single astrophysical model. Our technique combines SED model fitting with color-color and color-magnitude criteria to take advantage of all available IR photometric datapoints for each source (see also Povich et al. 2009; Povich & Whitney 2010; Smith et al. 2010; Povich et al. 2011).

### 3.1. Preparation of Photometry Tables for SED Fitting

We use a version of the least-squares, SED fitting tool of Robitaille et al. (2007, hereafter RW07) that batch-processes large numbers of sources using locally-compiled code. This is faster, more efficient, and more flexible compared to the more widely used RW07 web-based fitting tool.[1] To prepare our photometry for SED fitting, we performed the following steps:

1. We chose the provenance of our NIR photometry on a per-band, per-source basis. We preferred a high-quality (00 flag) UKIRT measurement wherever available, replacing the corresponding *2MASS* photometry where appropriate. In general, *2MASS* photometry was used for targets with no UKIRT observations, and otherwise for bright sources that saturated the UKIRT images or for UKIRT sources affected by artifacts. The filter response functions are applied directly to the SED models by the RW07 software (which does calculations in flux space), hence it was not necessary to shift the *2MASS* and UKIRT photometry into a common photometric system.

2. To mitigate the effects of unreported systematic uncertainties and intrinsic source variability,

---

[1] Go to http://caravan.astro.wisc.edu/protostars/.

Table 1.  Basic Data and Input Parameters Used for SED Fitting

|  | (l, b)[a] | IRAC Cat.[b] | UKIRT[c] (Y/N) | Stellar $A_{V,\mathrm{max}}$[d] (mag) | YSO Model Fits $A_{V,\mathrm{max}}$ (mag) | $d_{\mathrm{min}}$ (kpc) | $d_{\mathrm{max}}$ (kpc) |
|---|---|---|---|---|---|---|---|
| Flame Nebula | 206.5–16.4 | K13 | N | 40 | 80 | 0.41 | 0.42 |
| W40 | 28.8+03.5 | K13 | N | 40 | 80 | 0.4 | 0.7 |
| RCW 36 | 265.1+01.4 | K13 | N | 40 | 80 | 0.5 | 0.9 |
| NGC 2264 | 203.0+02.2 | K13 | Y | 30 | 60 | 0.90 | 0.93 |
| Rosette Nebula | 206.3–02.1 | K13 | Y | 40 | 80 | 1.2 | 1.4 |
| Lagoon Nebula | 6.0–01.2 | GII+3D | Y | 40 | 80 | 0.8 | 1.8 |
| NGC 2362 | 238.2–05.6 | K13 | Y | 5 | 5 | 1.4 | 1.6 |
| DR 21 | 81.7+00.5 | K13 | Y | 40 | 80 | 1.42 | 1.56 |
| RCW 38 | 268.0–01.0 | VC | N | 40 | 40 | 0.8 | 2.6 |
| NGC 6334 | 351.1+00.7 | GII | Y | 45 | 90 | 1.6 | 1.8 |
| NGC 6357 | 353.0+00.9 | GII | Y | 45 | 90 | 1.6 | 1.8 |
| Eagle Nebula | 17.0+00.8 | GI | Y | 45 | 90 | 1.6 | 1.8 |
| M17 | 15.1–00.7 | GI | Y | 40 | 80 | 1.9 | 2.1 |
| W3 | 133.9+01.1 | G | N | 40 | 80 | 1.9 | 2.1 |
| W4 | 134.7+00.9 | K13 | N | 40 | 40 | 1.9 | 2.1 |
| Trifid Nebula | 7.0–00.3 | GII | Y | 40 | 80 | 2.2 | 3.2 |
| NGC 3576 | 291.3–00.7 | VC | N | 40 | 40 | 2.7 | 2.9 |
| NGC 1893 | 173.6–01.7 | K13 | Y | 10 | 15 | 3.4 | 3.8 |

[a]Central location of MYStIX field (Feigelson et al. 2013).

[b]The IRAC photometry catalogs were obtained from the following sources: Kuhn et al. (2013; K13), GLIMPSE I (GI), GLIMPSE II (GII), GLIMPSE 3D (Lagoon only), the Vela-Carina Survey (VC; see Zasowski et al. 2009; Povich et al. 2011), and a custom run of the GLIMPSE pipeline (G; W3 only).

[c]UKIRT photometry, where available, came from King et al. (2013).

[d]For the MYStIX target regions with only 2MASS near-IR photometry available, we used a default value of $A_{V,\mathrm{max}} = 40$ mag for the reddened stellar photosphere fits.

Table 2. Source Tallies At Each Stage of MIRES Catalog Construction

|  | (1) IRAC Cat. | (2A)[a] Well-fit stellar Yes | (2B) No | (3A)[c] Marginal IRE Yes | (3B) No | (4A) Well-fit YSO Yes | (4B)[b] No | (5) Final MIRES |
|---|---|---|---|---|---|---|---|---|
| Flame Nebula | 18185 | 4616 | 1115 | 273 | 842 | 642 | 200 | 642 |
| W40 | 475903 | 100278 | 10171 | 5788 | 4383 | 4240 | 143 | 4240 |
| RCW 36 | 723 | 138 | 254 | 35 | 219 | 190 | 29 | 190 |
| NGC 2264 | 22363 | 16527 | 3184 | 1730 | 1454 | 1320 | 134 | 1330 |
| Rosette Nebula | 39079 | 34630 | 3039 | 1826 | 1213 | 1130 | 83 | 1135 |
| Lagoon Nebula | 157593 | 143608 | 9254 | 8064 | 1190 | 1106 | 84 | 1108 |
| NGC 2362 | 16396 | 11481 | 1959 | 800 | 1159 | 1065 | 94 | 1065 |
| DR 21 | 21727 | 12945 | 4633 | 3034 | 1599 | 1494 | 105 | 1498 |
| RCW 38 | 16019 | 13645 | 1361 | 640 | 721 | 717 | 4 | 717 |
| NGC 6334 | 134000 | 110235 | 15305 | 13728 | 1577 | 1190 | 366 | 1211 |
| NGC 6357 | 156664 | 126382 | 20204 | 18613 | 1591 | 1055 | 529 | 1062 |
| Eagle Nebula | 96768 | 85800 | 6691 | 5151 | 1540 | 1200 | 325 | 1215 |
| M17 | 215044 | 169205 | 26410 | 24570 | 1840 | 1137 | 703 | 1137 |
| W3 | 10733 | 4496 | 484 | 300 | 184 | 183 | 1 | 184 |
| W4 | 38540 | 9208 | 2434 | 1063 | 1371 | 1314 | 57 | 1314 |
| Trifid Nebula | 94029 | 73145 | 14897 | 14006 | 891 | 524 | 367 | 540 |
| NGC 3576 | 45879 | 39564 | 2292 | 1467 | 825 | 786 | 39 | 790 |
| NGC 1893 | 12401 | 7838 | 2236 | 726 | 1510 | 1340 | 169 | 1341 |
| Total | 1572046 | 963741 | 125923 | 101814 | 24109 | 20633 | 3432 | 20719 |

[a] All sources in columns 2A and 3A were discarded from the MIRES sample.

[b] SEDs of all sources in column 4B were visually reviewed. The number of sources in column 4B ultimately "rescued" for inclusion in MIRES can be found by subtracting column 4A from column 5 (this number is 86 total, but zero for certain fields).

– 10 –we set the *minimum* uncertainty used in SED fitting to 5% in the $JHK_S$, [3.6], and [4.5] bands and 10% for IRAC [5.8] and [8.0]. These floor values were used only where the reported uncertainties in the catalogs were smaller, otherwise the original uncertainties were used. It is important to note that these reset uncertainties are the ones published in MIRES, because they are used in all of our analysis; for *original* uncertainties we refer the reader to the source catalogs referenced in the previous section.

### 3.2. Filtering Out Non– and Marginal–IR-Excess Sources

As the first step in our filtering process, we fit reddened Castelli & Kurucz (2004) stellar atmosphere models, using the extinction law of Indebetouw et al. (2005), to all sources in our merged photometric catalogs that have $N_{\rm data} \geq 4$ detections among the 7 combined NIR–MIR bands. The reddening $A_V$, a free parameter in the RW07 fitting procedure, was allowed to range from 0 to a maximum value $A_{V,\rm max}$ determined independently for targets observed with UKIRT (Table 1) by inspection of sources plotted on a $J-H$ vs. $H-K_s$ color-color diagram. Sources for which the goodness-of-fit parameter $\chi_0^2$ for the best-fit model satisfied $\chi_0^2/N_{\rm data} \leq 2$ were considered well-fit by stellar photospheres and were removed from consideration for MIRES. The tallies of non-IRE versus possible IRE sources are given for each target in columns 2A and 2B of Table 2; note that the difference between the sum of these two columns and the values in column (1) is the number of sources with $N_{\rm data} < 4$.

Next we filtered out "marginal" IRE sources using the color cuts described by Povich et al. (2011) (see Appendix A for details of the color cuts). Sources classified as marginal IRE (column 3A in Table 2) are not considered for inclusion in MIRES because in general an excess appearing in a single IRAC band is consistent with systematics in the photometry, and is not strong evidence for youth (Smith et al. 2010; Povich et al. 2011). Marginal IRE sources tend to appear near the point-source detection limit in the [5.8] or [8.0] band and are more prevalent in areas of elevated nebulosity. The marginal IRE filter also captures objects with anomalously blue [5.8] − [8.0] colors or [3.6] − [4.5] colors consistent with interstellar reddening in the absence of longer-wavelength photometry. We estimate that 2% of marginal IRE sources are YSOs (see Appendix A). To illustrate the various data pathologies responsible for the large majority of marginal IRE source classifications, in Figure 1 we show example plots of SEDs with the best-fit stellar photosphere models. Most of these sources are relatively faint, falling near or below the detection limit at [5.8] or [8.0], as set by the local nebular background emission. Among the sources plotted in Figure 1, only G006.7218–00.2990 presents a plausible intrinsic IRE, but the non-detection of the source at [8.0] renders the apparent [5.8] excess suspicious.

The source populations passing through *both* of our initial filters (column 3B of Table 2) are dominated by *significant* IRE sources (but some are strongly variable stars or NIR–MIR catalog mismatches, which we deal with below). Our conclusion that the large majority of marginal IRE sources are not YSOs is supported by comparing the spatial distributions of sources rejected by this filter to those that pass it. In Figure 2 we show this comparison using the prototype MYStIX targets, NGC 2264 and the Trifid Nebula (Feigelson et al. 2013). We note that marginal IRE sources (magenta) are distributed quasi-uniformly throughout the fields while the significant IRE sources (cyan) are strongly clustered in the target MSFRs. In NGC 2264, which has deep IRAC data, the marginal IRE filter may also capture extragalactic sources. In the Trifid field, the distribution of marginal IRE sources is clearly non-uniform, reflecting large-scale spatial variations in interstellar reddening and nebular emission imprinted on the dense population of field stars in the inner Galaxy.

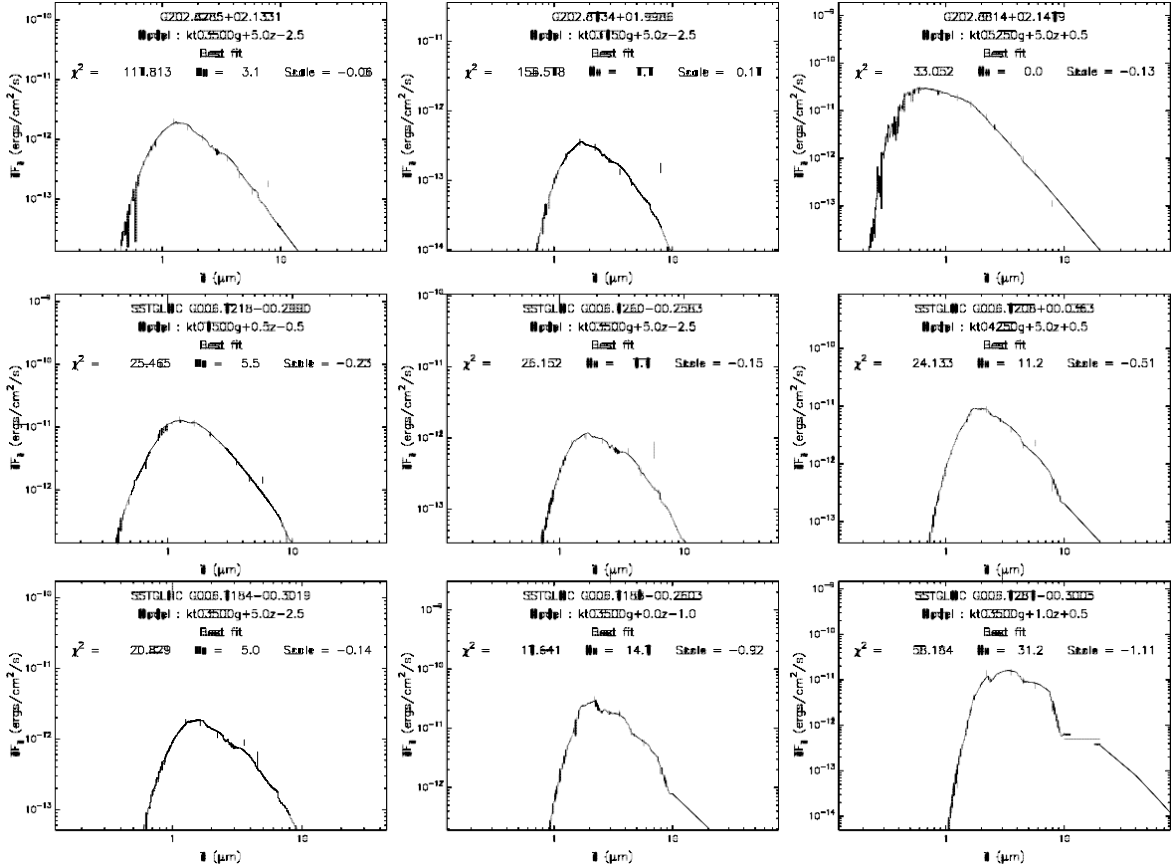

Fig. 1.— Example plots of SEDs (dots with error bars) classified as "marginal" IRE and best-fit stellar atmosphere models (curves). Each panel is labeled with the IRAC source designation and information about the best-fit SED: Castelli & Kurucz (2004) model designation, $\chi^2$, interstellar reddening ($A_V$ mag), and scale factor (log [$d$/kpc × $R$ /$R$]). *Top row:* Examples where the [8.0] band caused the stellar fit to fail, possibly due to poor [8.0] photometry (left and right panels, respectively) or spurious extraction of nebular contamination (center panel). *Middle row:* Examples where the [5.8] band (in the absence of an [8.0] detection) produced an apparently spurious excess, causing the fit to fail. *Bottom row:* Examples where two bands with color close to zero together caused the stellar fits to fail, due to variability, NIR/MIR mismatch (left panel; [3.6] − [4.5] ≈ 0) or very cool intrinsic photospheres with high reddening (center and right panels; [5.8] − [8.0] ≈ 0).

### 3.3. Fitting Significant IR-Excess SEDs with Young Stellar Object Models

After filtering out marginal IRE sources, we fit the SEDs of all remaining sources with star+disk+envelope radiation transfer models of young stellar objects (YSOs) from Robitaille et al. (2006, hereafter RW06). Sources for which $\chi_0^2/N_{\rm data} \leq 4$ were considered well-fit. The relaxation of the goodness-of-fit criterion compared to the previous case of fitting stellar atmospheres is necessary because (1) real YSOs are intrinsically variable sources, and our data come from multiple epochs, and (2) the RW06 model SEDs sparsely sample a very large parameter space. The tallies of sources with successful versus failed RW06 model fits for each target are given in columns 4A and 4B, respectively, of Table 2.

The RW06 models include only radiation from the central star and circumstellar dust, and therefore the model fits may give inaccurate results or fail entirely in the presence of significant emission lines originating in circumstellar gas. In particular, the [4.5] band can be brightened significantly by shocked molecular line emission produced by protostellar outflows (likely related to the "extended green object" phenomenon, see Cyganowski et al. 2008). G09 include a color cut to exclude "shock emission" from their IRE samples. This cut assumes such sources are unresolved knots in otherwise extended structures, an assumption that only works for nearby regions. All of the MYStIX MSFRs are more distant than the regions studied by G09, and the shocked emission is more likely to be confused with the driving YSO itself, as evidenced by the coincidence of point sources satisfying the G09 criteria with bright 24 $\mu$m point sources (Povich & Whitney 2010; Povich et al. 2011).

We adopt the strategy of Povich et al. (2011) for dealing with [4.5]-excess emission due to shocks, hybridizing their criteria for identifying shock emission with those of G09. Sources that satisfy both of the following criteria:

$$[3.6] - [4.5] > 1.1$$

$$[3.6] - [4.5] > \frac{1.9}{1.2} \times ([4.5] - [5.8]),$$

are labeled as candidate [4.5]-excess (hereafter [4.5]E) objects (see Fig. 3). Rather than discard these sources, we treated the [4.5] flux as an upper limit when fitting their SEDs with RW06 YSO models and assigned the flag value −99.99 to the photometric uncertainty on [4.5] (IRmag_err in MIRES; Table 3, column 5).

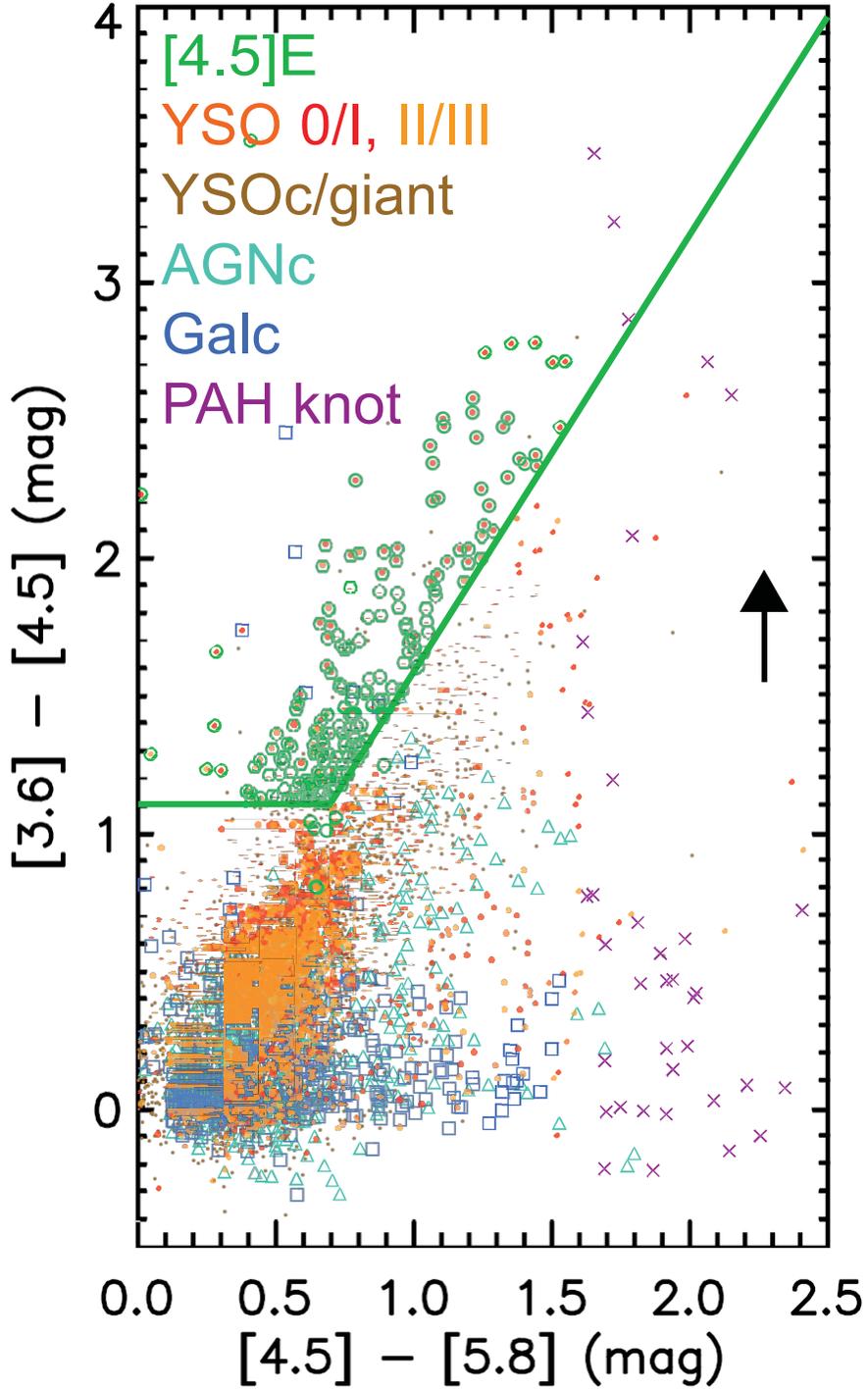

Fig. 3.— Color-color diagram illustrating the cuts used to identify YSOs (dots) affected by 4.5 μm excess ([4.5]E) emission likely due to shocked molecular lines (circled). All [4.5]E candidates are plotted, along with all other MIRES sources with photometric uncertainties ≤0.1 mag. The reddening vector corresponds to $A_V = 30$ mag. (A color version of this figure is available in the online Journal.)

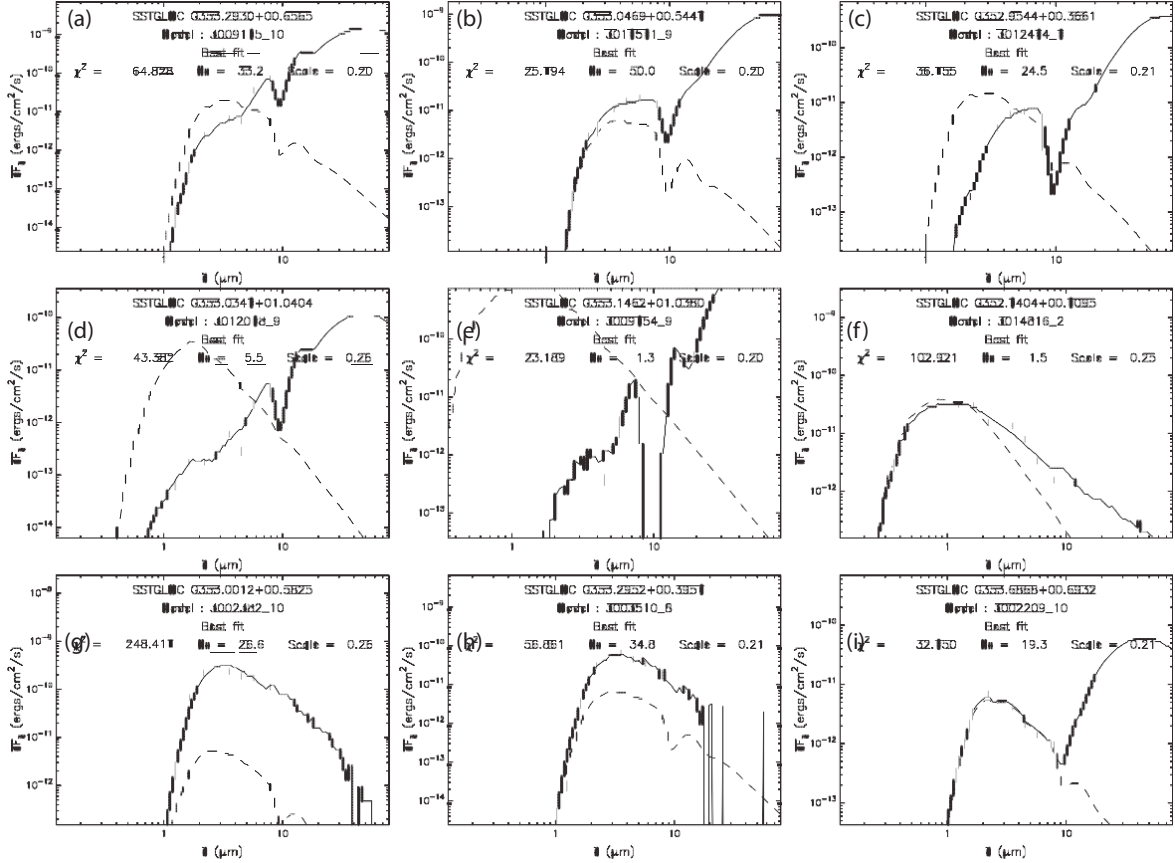

Fig. 4.— Example plots of SEDs (dots with error bars) that were poorly-fit with RW06 models (curves). Dashed curves show the stellar photospheres of the best-fit RW06 model YSOs as they would appear in the absence of circumstellar reddening from disks and/or envelopes. *Top row:* Examples of sources that were judged to be legitimate protostellar candidates during visual review and manually entered into MIRES. *Center and Bottom rows:* Examples of sources that were eliminated from consideration for MIRES after visual review (see text).

For the final step in constructing MIRES, we visually reviewed all SEDs and model fits for sources that were (1) poorly-fit by stellar atmospheres, (2) not classified as marginal IRE, and (3) were poorly-fit by RW06 models (i.e. the 3432 sources in column 4B of Table 2). This check was necessary to prevent our discarding interesting protostellar sources, in particular massive YSOs that can fail to be fit by RW06 models, for example due to excitation of polycyclic aromatic hydrocarbon (PAH) molecules in the disks or uncertainties in the interstellar extinction law. In many cases, we visually reviewed sources in the original MIR images and noted whether they were located in clusters, IR dark clouds, or bright-rimmed clouds, any of which increases the likelihood that a source is protostellar.

Figure 4 shows example SED plots, drawn from the NGC 6357 field, illustrating various common pathologies for sources poorly fit by either stellar atmospheres or RW06 YSO models. The top row of panels shows examples of "rescues" judged by eye to be legitimate candidates for MIRES: (a) An SED revealing a composite cool dust continuum plus an intrinsic PAH emission spectrum shows monotonic increase in flux density from $J$ to [8.0], with the exception of a dip at [4.5] as this is the only IRAC band that does not contain a PAH emission feature. (b) A potentially variable protostar shows a very red SED, but the $K_S$ flux density is elevated above the model while the [3.6] flux density falls below, causing a poor fit. (c) A likely protostar barely missed the $\chi^2/N_{\rm data} \leq 4$ cut because of a [8.0] detection due to strong silicate absorption at 9.7 $\mu$m. The number of rescued sources for each target (which can be zero) can be found by subtracting column 4A from column 5 in Table 2. The middle and bottom panel rows show common examples MIRES "rejects" that were poorly-fit by RW06 models: (d) Match of a faint star with PAH nebular contamination in IRAC bands and (e) PAH nebular knot both show a characteristic "check-mark" morphology in the IRAC SED, in which the [4.5] band is sharply suppressed compared to the other 3 bands. (f) A strongly variable star or an NIR–MIR mismatch produces a "broken" SED that otherwise resembles a normally-reddened stellar photosphere. (g, h) Likely asymptotic giant branch (AGB) stars with dust-rich winds (a category that includes carbon stars and OH/IR stars) have SEDs characterized by a precipitous rise with increasing wavelength through the NIR bands followed by a flattening/decline with very bright ($\lambda F_\lambda$ $10^{-10}$ erg cm$^2$ s$^{-1}$) emission through the IRAC bands. (i) A cool, field giant barely missed the $\chi^2$ cut for well-fit by reddened photosphere and does not show significant IRE above the photosphere of the central star in the best-fit RW06 model.

The locations of the sources shown in Figure 4 along with all MIRES in the NGC 6357 field (Table 2, column 5) and other sources rejected on the basis of failed RW06 model fits (Table 2, column 4B) are overlaid on an 8.0 $\mu$m image in Figure 5. The spatial distributions of these different source populations illustrate both the effectiveness of using the RW06 model fits as a filter and the need for a final visual review. While the MIRES that are well-fit with RW06 YSO models populate several large clusters and a more distributed component to the population, the poorly-fit sources are found exclusively in a distributed mode, biased away from the central clusters. Only seven of the final MIRES originated as SEDs poorly-fit by RW06 models and subsequently "rescued" by the visual review, far too few to impact significantly the global spatial distributions. We note that the three example rescues highlighted in yellow are all found in or near real YSO clusters, IR dark clouds, or bright-rimmed clouds.

## 4. SED Fitting Results and Analysis

All IRE sources passing through the filtering process described in the previous section are entered into MIRES, which is available as a single machine-readable table in the online edition of the Journal. Table 3 describes the columns in MIRES. Columns (1) through (8) are basic IR source properties and matching results from the combined MIR and NIR source catalogs (GLIMPSE, K13, and King et al. 2013). In this section we describe columns (9) through (12), which give the basic SED fitting results.

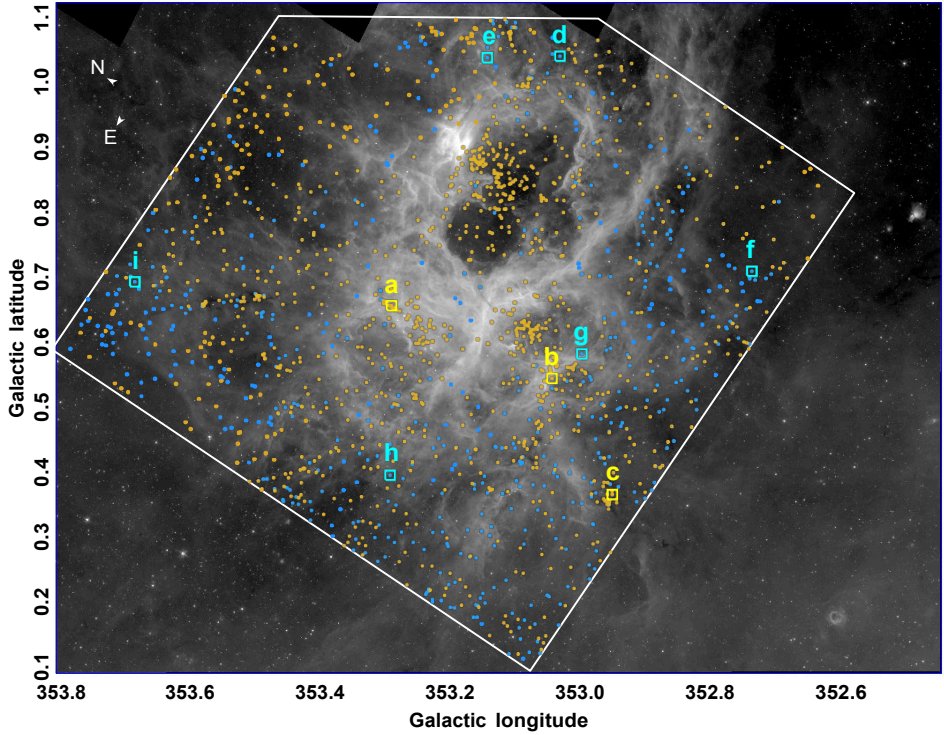

Fig. 5.— GLIMPSE 8 $\mu$m mosaic of NGC 6357, with positions of sources included in MIRES (orange) and those poorly-fit (blue) by RW06 YSO models and excluded from MIRES overlaid. Locations of the sources providing the example SEDs shown in Fig. 4 are marked by boxes and labeled by panel letter.

### 4.1. $A_V$ from the SED fits

Following Povich et al. (2011) and previous work, for each MIRES SED we define the set $i$ of well-fit RW06 YSO models according to

$$\frac{\chi_i^2 - \chi_0^2}{N_{\text{data}}} \leq 2 \qquad (1)$$

where $\chi_0^2/N_{\text{data}}$ is the data-normalized goodness-of-fit parameter for the best-fit model (Column 10 of Table 3). We then assign a $\chi^2$–weighted probability to each model using

$$P_i = P_n\, e^{-\chi_i^2/2}, \qquad (2)$$

Table 3. MIRES Catalog Format

|  | Column Label | Description |
|---|---|---|
| (1) | MIR_Name | Source name in IRAC Catalog (GLIMPSE or K13) |
| (2) | RAdeg | Right ascension (J2000, degrees) |
| (3) | Dedeg | Declination (J2000, degrees) |
| (4) | IRmag | Magnitudes in 7 IR bands used for SED fitting: $J$, $H$, $K_S$, [3.6], [4.5], [5.8], [8.0] |
| (5) | IRmag_err | Uncertainties on the 7 IR magnitudes used for SED fitting,[a] *reset to floor values*[b] |
| (6) | NIRphot_cat | Provenance of *each* of 3 $JHK_S$ sources matched to IRAC source: 0=*2MASS*, 1=UKIRT, −1=none |
| (7) | UKIDSS_label | UKIRT catalog source name (King et al. 2013) matched to IRAC source |
| (8) | NIRphot_num_SM | Number of UKIRT sources providing possible alternative matches to IRAC source |
| (9) | SED_flg | Source classification flag: 0=likely YSO, 1=starburst galaxy, 2=AGN, 3=PAH knot |
| (10) | SED_chisq_norm | $\chi_0^2/N_{\rm data}$ of best-fit SED model.[c] |
| (11) | SED_AV | Visual extinction $A_V$ determined from $\chi^2$–weighted mean of all acceptable SED fits |
| (12) | SED_stage | Evolutionary Stage classification, RW06 YSO models: 1=Stage 0/I, 2=Stage II/III, −1=ambiguous[d] |
| (13) | Prob_dens | $= 1 - f_{\rm con}$, where $f_{\rm con}$ is the fraction of MIRES in the local neighborhood that are consistent with foreground/background contaminants.[e] |
| (14) | MEM_flg | =1 if probable member of target MSFR, 0 otherwise |
| (15) | XFOV | =1 if source falls within *Chandra* field-of-view for MYStIX, 0 otherwise |
| (16) | MYStIX_SFR | Name of MYStIX MSFR |

[a] Value of −99.99 means the corresponding flux measurement was used as an upper limit for SED fitting.

[b] As described in Section 3.1, minimum uncertainty used for SED fitting was set in flux density space using $\delta F_i \geq 0.05 F_i$ for $JHK_S$, [3.6], and [4.5] and $\delta F_i \geq 0.10 F_i$ for [5.8] and [8.0]. For *original* photometric error bars we refer the reader to the appropriate original source catalogs: K13, King et al. (2013), or GLIMPSE.

[c] For [4.5]E SEDs where [4.5] flux was used as an upper limit, $3 \leq N_{\rm data} \leq 6$, for all other SEDs $4 \leq N_{\rm data} \leq 7$.

[d] All sources in MIRES, regardless of SED_flg, were fit with RW06 models and hence can be classified according to YSO evolutionary Stage.

[e] NaN values are assigned to MIRES catalog sources falling within designated "control" fields for clustering analysis.

where $P_n$ is chosen such that $P_i = 1$. We can use the resulting probability distributions of model parameters to characterize and constrain key physical and observational parameters of each source.

A useful parameter is the interstellar reddening $A_V$, expressed in magnitudes of V-band extinction between the observer and the outer edge of the disk or protostellar envelope in the RW06 models. We compute the weighted-mean interstellar extinction based on the SED fits to each MIRES as

$$A_{V,\mathrm{SED}} = \sum_i P_i A_{V,i} \tag{3}$$

(Column 11 of Table 3). We use $A_{V,\mathrm{SED}}$ to help distinguish between faint YSOs and likely extragalactic contaminants (see Section 4.2 below). We caution, however, that in some cases the SED modeling cannot constrain the $A_V$ parameter, and $A_{V,\mathrm{SED}}$ defaults to $A_{V,\mathrm{max}}/2$ (Table 1). In particular, we expect the $A_V$ parameter to be poorly constrained for:

1. Target regions with low obscuration. If the actual interstellar reddening is low ($A_V \lesssim 5$ mag), the effect on IR SEDs of intrinsically red sources is small and difficult to measure. For example, the NGC 2362 cluster has completely dispersed its natal cloud and therefore has reddening near zero, but its MIRES catalog entries tend toward $A_{V,\mathrm{SED}} \sim 2.5$ mag, as we allowed for up to $A_{V,\mathrm{max}} = 5$ mag of reddening in the SED model fitting.

2. Sources missing $J$ and $H$ photometry measurements. The blue end of our SEDs is most affected by reddening, hence if NIR datapoints are missing we do not expect to achieve good constraints on $A_V$. We recommend that any future investigations of interstellar reddening based on MIRES be restricted to the subset of sources with reported detections at $H$ at minimum, and preferably both $J$ and $H$ detections.

3. YSOs obscured by nearly edge-on disks or deeply embedded in protostellar envelopes. In the cases of the reddest MIRES, the SED (including the NIR datapoints, if present) is likely dominated by emission/absorption from the disk and envelope, which completely veil the central star. In such cases (generally Stage 0/I and Ambiguous YSOs, see below) the $A_{V,\mathrm{SED}}$ values should be viewed with suspicion.

### 4.2. Flagging Candidate Starburst Galaxies, AGN, and PAH Nebular Knots

Several types of intrinsically red, contaminating objects in the MIRES catalog can masquerade as YSOs. Unresolved extragalactic sources dominate the faint, red source populations in deep *Spitzer* observations of nearby star-forming clouds at high Galactic latitude or more distant MSFRs on sightlines toward the outer Galaxy (G09, Beerer et al. 2010). For inner Galaxy MSFRs observed as part of the shallower GLIMPSE or Vela–Carina surveys (see Table 1), extragalactic contaminants are not expected in significant numbers (Kang et al. 2009), but for consistency in building MIRES we apply the same procedure for flagging contaminants to all target regions. H II regions with bright nebular PAH emission present a different type of contamination, unresolved nebular knots that can either appear as "spurious," very red point sources in their own right (K13) or contaminate the extraction apertures of stars detected at shorter wavelengths, producing apparent excess emission at longer wavelengths (G09). While some extragalactic sources or PAH nebular contamination may be captured by the marginal IRE filter, sources with strong excess in multiple bands will not be filtered out.

We use the IR color spaces shown in Figure 6 in conjunction with ancillary information from the SED fitting and color-magnitude information to flag candidate extragalactic sources and sources affected by PAH nebular contamination. We flag such sources using SED_flg (Column 9 of Table 3) rather than remove them from MIRES because additional photometric or spectroscopic data in the future could confirm that some are indeed young stellar members of the target MSFRs. For candidate YSOs (YSOc), SED_flg = 0. Our membership analysis finds a (small) fraction of probable members with SED_flg = 0 (see Section 5 below).

Extragalactic contaminants come in two flavors, starburst galaxies (Galc; SED_flg = 1) with strong PAH emission enhancing the [8.0] band, and obscured active galactic nuclei (AGNc; SED_flg = 2) with intrinsic dust emission (G09). Starburst galaxies are found in the lower-right portions of the color-color diagrams shown in Figures 6$a$ and $c$; the color cuts used by G09 to identify Galc are plotted as thick black boundary lines. AGNc do not separate cleanly from YSOc in color space, so G09 define a cut in [4.5] versus [4.5] − [8.0] color-magnitude space to identify AGNc, which tend to be faint ($[4.5]_0 > 13.5$ mag).

The G09 color-color and color-magnitude cuts were based on the loci of extragalactic sources detected in deep *Spitzer*/IRAC observations of fields that contained neither significant numbers of Galactic point sources nor foreground reddening. Thus to avoid mis-classifying legitimate YSO members of Galactic MSFRs, we first deredden each source using the extinction law of Indebetouw et al. (2005) scaled to its $A_{V,\mathrm{SED}}$ (Table 3, Column 11), and then apply the G09 cuts (we note that the cut shown in Fig. 6$a$ is essentially reddening-free). The MIRES catalog also includes faint sources that were not detected in the [8.0] band and hence cannot be evaluated against the G09 cuts, so we flag all such sources with dereddened $[3.6]_0 > 14.5$ mag as Galc (SED_flg = 1).

Because the magnitude distribution of the extragalactic background is a function of reddening only, while that of the YSO population in a given Galactic MSFR is a function of both reddening and distance, the degree to which the low-luminosity tail of a YSO population overlaps with the extragalactic contaminating population (particularly AGNc) increases with increasing distance, and is worse for regions with no significant intervening reddening column behind the YSOs but in front of the extragalactic background. For these reasons, it can be easy to mis-classify a legitimate YSO as an extragalactic contaminant, so the Galc and AGNc flags do not automatically disqualify MIRES for membership in a MYStIX MSFR (see Section 5.1). We discuss the impact of extragalactic contaminants on individual MSFRs in Appendix B; here we note that, as expected, we find significant numbers of extragalactic contaminants in the MYStIX targets with deeper *Spitzer*/IRAC observa-tions (K13; see Table 1), and negligible numbers in targets with shallower, GLIMPSE observations.

Sources were flagged as PAH nebular knots (SED_flg = 3; Table 3, Column 11) and *rejected* for further consideration for membership if they satisfied *both* of the following criteria (Fig. 6*a*):

$$[4.5] - [5.8] \geq 1.6$$

$$[5.8] - [8.0] \geq 0.5.$$

The above criteria may also select massive YSOs with sufficient UV radiation to excite PAH emission in their own disks. Massive YSOs may be distinguished by very red continuum emission in the $K_s - [4.5]$ color, which is free from PAH contamination, hence we *excluded* such objects from the PAH nebular knot flag using

$$K_s - [4.5] > [4.5] - [5.8]$$

(Fig. 6*b*). We note that G09 presented their own scheme for flagging "PAH aperture contamination" that is similar to ours in some respects, but it is more aggressive in selecting sources with bluer $[3.6] - [4.5]$ and $[4.5] - [5.8]$ colors. Whitney et al. (2003) found that protostars can be blue at $[3.6] - [4.5]$ thanks to scattered light off of envelope cavity inner walls. G09 studied a set of nearby, low-mass star-forming clouds that included few massive YSOs, while the more distance MYStIX target regions include many more luminous YSOs with the potential for intrinsic PAH excitation in their disks/envelopes.

The large majority of PAH nebular knots and potential extragalactic sources in MIRES came from *Spitzer* data processed by K13 rather than from GLIMPSE, in part because the observations went deeper, but also because, compared to the GLIMPSE pipeline, their point-source detection and aperture photometry extraction were less conservative in rejecting marginally resolved sources. The majority of Galc, AGNc, and PAH knots have relatively high MIR photometric uncertainty (K13), which is consistent with their faintness but also supports the idea that many are marginally resolved. In Figure 6 we show all MIRES with photometric uncertainties ≤0.2 mag in the relevant bands, including 1761 Galc, 1920 AGNc, and 168 PAH knots. If instead we plotted (in panels *a* and *c*) only sources with uncertainties ≤0.1 mag (as in the other color-color diagrams presented in this work), 261 Galc, 360 AGNc, and 35 PAH knots would remain.

### 4.3. Evolutionary Stage from the SED fits

The RW06 YSO models can be divided into evolutionary stages that parallel the well-known empirical T Tauri classification scheme: MIR emission from Stage 0/I YSOs is dominated by infalling, dusty envelopes; Stage II and III YSOs are dominated by optically thick and optically thin circumstellar disks, respectively. Following previous work (Povich et al. 2009; Povich & Whitney 2010; Smith et al. 2010; Povich et al. 2011) we use Equation 3 to compute the probability distribution of evolutionary stage for each MIRES and classify each as Stage 0/I or II/III if $P_i \geq 0.67$; if this criterion is not met the stage is considered "Ambiguous." Examples of the best-fit models to the SEDs of MIRES representing three different target MSFRs and the full range of evolutionary stage classifications are shown in Figure 7. The stage classifications are recorded in the SED_stage column of MIRES (Column 12 of Table 3).

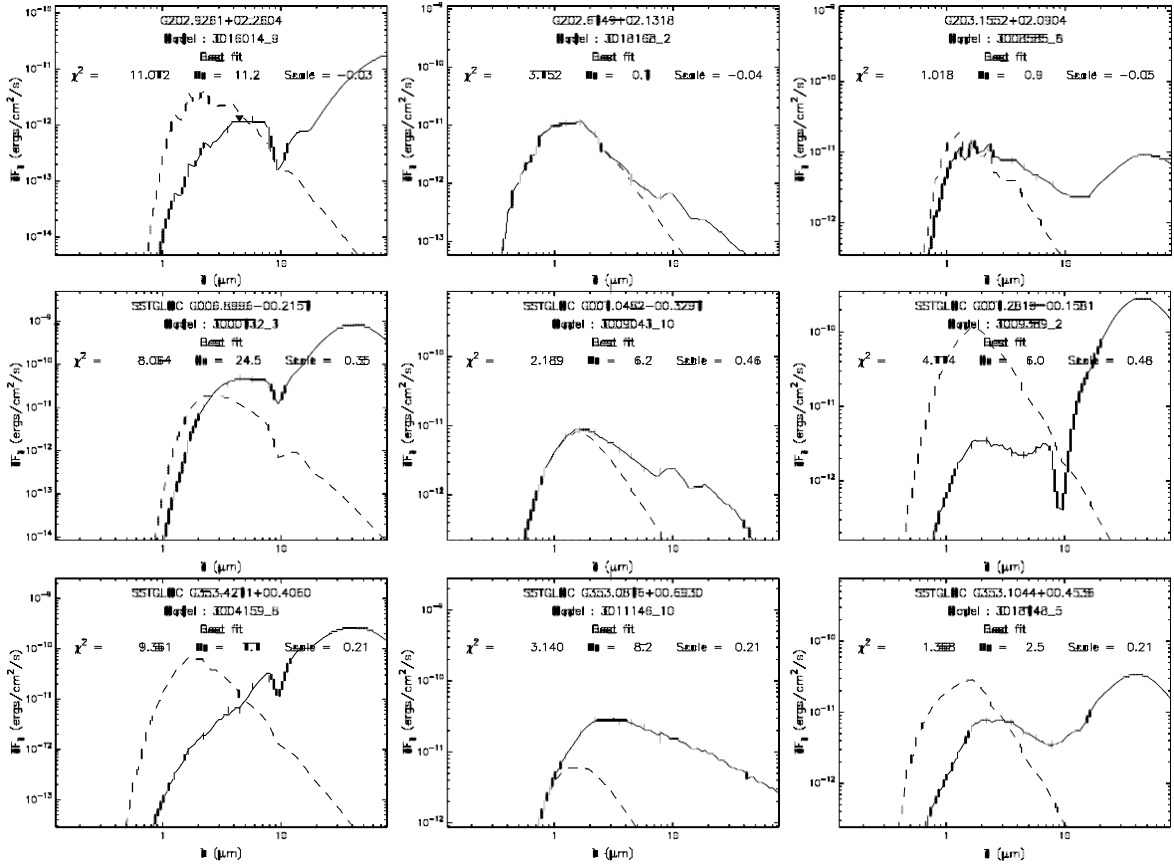

Fig. 7.— Example plots of SEDs (dots with error bars) that were will-fit with RW06 models (curves). Dashed curves and annotations are the same as in Fig. 4. *Panel rows* correspond to sources in different target MSFRs: NGC 2264 (*top*), Trifid (*center*), and NGC 6357 (*bottom*). *Panel columns* correspond to sources with different most probable evolutionary stage: 0/I (*left*), II/III (*center*), and ambiguous (*right*).

While MIRES incorporates only photometric data from 1–8 $\mu$m, the intrinsic SED emission from a cool, infalling envelope peaks from 50–100 $\mu$m (RW06, and see also the left and right panel columns of Figure 7). We therefore have no constraints from our data in the thermal IR, where the difference between a disk-dominated and envelope-dominated SED is easiest to measure (Indebetouw et al. 2007). Our stage classifications from SED fitting are therefore based on extrapolating redward from the available data. As a consequence, Stage II objects with high interstellar reddening may be misclassified as Stage 0/I, and the fraction of MIRES with Ambiguous stage classifications is relatively high, 28%. Other approaches for classifying YSOs based on spectral indices or colors (e.g. G09) can be applied to MIRES using the photometry values given in Column 5 of Table 3, but we emphasize that *all* classification schemes based solely on some combination of $JH K_S$ and *Spitzer*/IRAC photometric data are vulnerable to the same extrapolation uncertainties. For future studies based on the analysis of evolutionary stage or class, we recommend that longer-wavelength photometry from available *Spitzer*/MIPS, *Herschel*, and/or *WISE* observations be used in conjunction with MIRES.

## 5. Identification of Probable Complex Members

The MIRES catalog includes both young stellar members of the MSFRs and a variety of unassociated contaminants. In addition to extragalactic sources and PAH nebular knots (see Section 4.2 and Table 3), dusty AGB stars, field giants with high interstellar reddening, and even unassociated YSOs may all be mistaken for YSO members. The last three categories of stellar contaminants, most prevalent in the inner Galactic plane fields covered by the various GLIMPSE Catalogs, cannot be readily distinguished from members using 1–8 $\mu$m photometry alone (Povich et al. 2009). Reddened giants and dusty AGB stars could be distinguished by matching MIRES with longer-wavelength photometry data from MIPS, *Herschel*, and/or *WISE* observations, as evolved stars have bluer $[8.0] - [24]$ colors compared to YSOs (see Povich et al. 2011, and references therein); this is beyond the scope of the present work.

In this section we describe the remaining four columns in the MIRES catalog (Table 3), which report the results of our membership analysis. Column 13 (Prob_dens) is a parameter related to the probability that a source at a given coordinate is a complex member, based on spatial clustering analysis, column 14 (MEM_flg) is a binary flag set if a MIRES is a probable member of the parent MSFR identified in column 16 (MYStIX_SFR). Column 15 (XFOV) is another binary flag set if the source is found within the field-of-view of the *Chandra*/ACIS observations used for MYStIX, which is a subset of the MIRES extended membership field. MIRES with both the MEM_flg and XFOV bits set are adopted as MPCMs (Broos et al. 2013).

## 5.1. Estimating Membership Probabilities from Spatial Distribution

We can leverage information about the spatial distribution of MIRES in and around each target MSFR to aid in the identification of YSO members. Members are expected to be spatially concentrated or "clustered" toward their parent MSFR, while contaminants, with the exception of PAH nebular knots and unassociated YSOs, should be uniformly distributed. We can therefore consider the spatial distribution of MIRES in a given target field to be a statistical mixture of clustered and distributed components. Each MIRES has a probability of association with either component, depending on its position (coordinates; Columns 2 and 3 of Table 3).

To establish a baseline for the surface density $\Sigma_{con}$ of contaminants (the distributed component), where possible we define "control fields" within the larger MIRES field around each target MSFR, selected to avoid both bright PAH nebulosity in the [8.0] images and evident clusters of MIRES. In the process we also defined a "primary target field" for each MSFR, which included the *Chandra* field-of-view (the "MYStIX field") plus any spatially contiguous regions containing clustered MIRES and/or molecular clouds as evidenced by bright or dark 8 $\mu$m diffuse emission (the "MIRES extended membership field"). The selection of target and control fields as applied to the Trifid Nebula and NGC 6357 is illustrated in Figure 8. Some MIRES fields contain other, unassociated young star clusters or star-forming clouds (for example, the NGC 3576 field also contains the famous massive young cluster NGC 3603 in the background), and in these cases we defined secondary target fields that were excluded from both the control and MIRES extended membership fields. For some MSFRs there was insufficient NIR/MIR coverage to establish a control field, so the spatial distribution analysis was omitted, as noted in Section 5 below.

While $\Sigma_{con}$ can be measured directly in the control fields, it may vary by a factor of a few within the target fields due to variations in sensitivity that can arise from background nebulosity, extinction, or source crowding. To estimate surface density of contaminants in the target fields, we assume that the sensitivity variations affect both MIRES contaminants and the far more numerous, non-IRE field stars (Column 2A of Table 2) similarly within a given MIRES field. We then use the surface density of field stars as a proxy for the spatial distribution of the unclustered (contaminant) component. Surface densities for all MIRES ($\Sigma$) and for non-IRE field stars ($\Sigma_{field}$) are calculated independently using kernel density estimation with a $\sigma = 1$ Gaussian kernel. The surface density of MIRES members as a function of position is then calculated as

$$\Sigma_{mem} = \Sigma - \Sigma_{field} \left. \frac{N}{N_{field}} \right|_{con} = \Sigma - \Sigma_{con}, \tag{4}$$

where $\Sigma_{con}$ and $\Sigma_{field}$ apply to the *full* MIRES field, and the scale factor between them is the ratio of MIRES to non-IRE field stars *in the control field*. The probability that any specific MIRES belongs to the clustered spatial component is then calculated as

$$\frac{\Sigma_{mem}}{\Sigma} \equiv 1 - f_{con}, \tag{5}$$

and reported in Column 13 of Table 3. The "contamination fraction," $f_{con}$, is the fraction of MIRES

that are expected to belong to the distributed component, as a function of position. The positional variation of $f_{con}$ in the Trifid and NGC 6357 fields is illustrated in Figure 8. The contamination fraction is indeed lowest in the dense, central clusters, increases toward the outer regions of each MSFR, and is undefined in the control fields. The results of the above membership probability calculations were visually reviewed for each MIRES field, and if evident clusters were found to have high $f_{con}$, or conversely if regions of low $f_{con}$ were found to extend to the boundary of the target field, the spatial boundaries of the target and control field were adjusted and the procedure was repeated iteratively.

For the purposes of identifying probable members, the various types of MIRES (as denoted by SED_flg, column 9 of Table 3) must be handled differently:

### 5.2. Candidate YSOs (YSOc Flags)

The large majority of MIRES members in any target MSFR are classified YSOc (SED_flg = 0), but the converse, that the majority of YSOc are members, need not be true in a given field. Figure 9 shows mid-IR images of the prototype MYStIX regions NGC 2264 and the Trifid Nebula. (Fig. 9, including additional panels for all 16 other MYStIX regions included in MIRES are available in the online Journal, is provided as a large figure set at the end of Appendix B.) The top panels of each figure pair show the spatial distributions of all MIRES, color-coded according to SED_flg, overlaid on IRAC [3.6] mosaic images of each field.

In the NGC 2264 field, YSOc are found almost exclusively in the clustered component, while the distributed component is dominated by extragalactic sources (Fig. 9$d$). NGC 2264 is located along a sightline toward the outer Galaxy, with relatively few field stars apparent in the [3.6] image. We therefore classify *all* MIRES flagged YSOc as probable members of NGC 2264.

In the Trifid field, by contrast, both the clustered and distributed components are dominated by YSOc (Fig. 9$p$). It is not immediately clear which component contains the greater number of sources, and it would definitely not be prudent to assume that all YSOc in this field are associated with the Trifid complex. Trifid is located in the inner Galaxy, and consequently the [3.6] image is dominated by field stars that contribute a significant fraction of contaminating, stellar sources to MIRES. For the case of Trifid, we therefore use $1 - f_{con}$ (column 13 of Table 3, see Section 5.1 above) as the probability that a given MIRES is part of the spatially clustered component, and define a threshold value max($f_{con}$) below which MIRES with YSOc flags are classified as probable members (MEM_flg = 1). To find a reasonable threshold, we first specify a *global* contamination fraction that we are willing to accept for the membership sample, $F_{con} = 0.15$ in the case of Trifid. We then compute max($f_{con}$) as follows:

1. Define the initial value max($f_{con}$) = $F_{con}$.

2. Define the subset $F = \{f_{con} : f_{con} \leq \max(f_{con})\}$ of MIRES spatially restricted to the primary

target field (delineated by green and blue boundary lines in Fig. 9*p*, top panel). $F$ thus represents MIRES nominated for probable membership.

3. Add sources iteratively to $F$ by increasing max($f_{con}$). Stop at the final, cutoff value of max($f_{con}$) when the (increasing) moving average of $F$ reaches or exceeds $F_{con}$.

The 18 MIRES target MSFRs are divided evenly between those resembling NGC 2264, with negligible contamination from unassociated stellar sources, and those resembling Trifid, with significant contamination among YSOc. The $F_{con}$ column under "YSOc Flags" in Table 4 identifies which MIRES fields, including Trifid, required spatial clustering analysis to establish YSOc membership; for the remaining fields, including NGC 2264, *all* MIRES flagged as YSOc (SED_flg = 0) were also flagged as members, so $F_{con} = 0$ by construction. Trifid represents a "worst-case" for YSOc contamination (joined by the Lagoon Nebula, M17, and NGC 3576), and for other MSFRs we were able to reduce $F_{con}$.

### 5.3. Candidate Extragalactic Point Sources (Galc/AGNc Flags)

MIRES flagged as extragalactic (Galc or AGNc with SED_flg = 1 or 2) are dominated by contaminants, but our inability to cleanly separate extragalactic sources from low-luminosity YSOs becomes evident in several MIRES fields, notably NGC 2264 and NGC 1893. NGC 1893 is the most distant target MSFR (Table 1), and it also was observed with a deep *Spitzer*/IRAC integration (K13). MIRES with extragalactic flags clearly cluster together with YSOc in the central regions of this field (Fig. 9*r*), meaning that they are most probably YSO members with apparent magnitudes falling in the range occupied by extragalactic background sources. In NGC 2264 we observe a weak tendency for extragalactic-flagged MIRES to cluster with the two dense YSOc clusters (Fig. 9*d*), some of these are also likely members.

For NGC 2264 and NGC 1893, in addition to the 9 other MIRES target fields for which we applied spatial clustering analysis to determine YSOc probable members, we define a new (more stringent) threshold

$$\max(f_{con}) = \max(f_{con}) - \left\langle \frac{N_{exgal}}{N} \right\rangle_F. \quad (6)$$

The second term is the fraction $N_{exgal}/N$ of MIRES in the subset $F$ of the primary target field that have both extragalactic flags (Galc or AGNc) and $f_{con} \leq \max(f_{con})$. MIRES with extragalactic flags and $f_{con} < \max(f_{con})$ are also classified as probable members (MEM_flg = 1). The columns under "Galc/AGNc Flags" in Table 4 give the fraction $F_{exgal}$ of probable members flagged as extragalactic (identically zero in the fields where no Galc/AGNc sources were considered for membership), and the threshold max($f_{con}$). With the exception of NGC 1893, $F_{exgal}$ is very small, 1%. The membership flag should be given priority over the extragalactic flags for subsequent population studies based on MIRES.

Table 4. Parameters Used to Identify Probable MIRES Members from Spatial Distributions

| | YSOc Flags | | Galc/AGNc Flags | |
|---|---|---|---|---|
| | $F_{con}$ | max ($f_{con}$) | $F_{exgal}$ | max ($f_{con}$) |
| Flame Nebula | 0 | ⋯ | 0 | ⋯ |
| W40 | 0.11 | 0.282 | 0.011 | 0.041 |
| RCW 36 | 0 | ⋯ | 0 | ⋯ |
| NGC 2264 | 0 | ⋯ | 0.017 | 0.027 |
| Rosette Nebula | 0 | ⋯ | 0 | ⋯ |
| Lagoon Nebula | 0.15 | 0.297 | 0.028 | 0.266 |
| NGC 2362 | 0 | ⋯ | 0 | ⋯ |
| DR 21 | 0 | ⋯ | 0 | ⋯ |
| RCW 38 | 0.07 | 0.223 | 0.011 | 0.210 |
| NGC 6334 | 0.11 | 0.299 | 0.009 | 0.290 |
| NGC 6357 | 0.10 | 0.219 | 0.002 | 0.217 |
| Eagle Nebula | 0.10 | 0.263 | 0.012 | 0.251 |
| M17 | 0.15 | 0.317 | 0.014 | 0.285 |
| W3 | 0 | ⋯ | 0 | ⋯ |
| W4 | 0 | ⋯ | 0 | ⋯ |
| Trifid Nebula | 0.15 | 0.376 | 0.024 | 0.349 |
| NGC 3576 | 0.15 | 0.261 | 0.023 | 0.220 |
| NGC 1893 | 0 | ⋯ | 0.211 | 0.279 |

### 5.4. PAH Nebular Knots

Among all the possible contaminating source populations, PAH nebular knots (SED_flg = 3) are the most pernicious. Because they appear in regions of bright nebular emission, their spatial distribution is highly non-uniform, with a tendency to cluster in the same locations as the young stellar populations responsible for producing the numerous bright H II regions targeted by MYStIX (see the Flame Nebula and W40, Fig. 9*a* and *b*, for two of the worst cases). We therefore reject any MIRES flagged as PAH nebular knots from consideration as probable MIRES complex members.

### 6. Discussion and Summary

Final tallies of MIRES, broken down by target MPCM and source classifications, are presented in Table 5. Probable members are drawn from the combined MYStIX X-ray fields and the MIRES extended membership fields and comprise 8686/20,719 = 41.9% of the IRAC Catalog. Anticipating that much future MYStIX-based science will concentrate on the combined X-ray and MIRES MPCM samples (Broos et al. 2013) that are spatially restricted to the MYStIX X-ray fields, in Table 6 we give final tallies for the subset of 8127/20,719 = 39.2% of MIRES located within the boundaries of the MYStIX fields. Within the MYStIX fields, the fraction of MIRES members is 5103/8127 = 62.8%, significantly higher than for MIRES as a whole. This implies that any residual contamination among the MIRES classified as members in the MYStIX fields is low (compared to the global contamination fractions $F_{\rm con}$ reported in Table 4), as expected given that the archival *Chandra*/ACIS observations used for MYStIX typically targeted dense, young stellar clusters.

A majority of MIRES probable members, 5103/8686 = 58.7%, are found within the MYStIX fields. This fraction is a lower limit to the true spatially-limited fraction of star formation activity in each MYStIX target sampled by the *Chandra* data, because the MYStIX fields generally contain the densest stellar clusters and brightest diffuse nebular emission (see Figs. 9) within the larger MIRES fields. Crowding and nebulosity conspire to reduce IR point-source sensitivity, while the contamination fraction among MIRES members is generally higher in the extended fields. Conversely, a significant minority of MIRES are located within the extended membership fields, and these often reveal regions of active star-formation associated with the MYStIX MSFRs that fall outside the *Chandra* FOV and hence were excluded from the MPCM tables provided by Broos et al. (2013).

As is the case for the MYStIX project as a whole (Feigelson et al. 2013), MIRES in no way provides a complete sample of the young stellar population within a given MSFR. The principal selection criteria for MIRES are detection of a point source in at least 4 of the 7 combined $JHK_S$ and IRAC bands, two of which must be IRAC [3.6] and [4.5], and measurement of a significant IRE from the available photometry. The effective depth of the IRAC Catalogs varies strongly among MIRES targets, due to differences in integration time, distance to the target stellar population, nebular background emission, and crowding in the field. Some target MSFRs have deep NIR

Table 5.  MIRES Catalog Tallies I: Full Fields

| MIRES Field | (1) All | (1A) S0/I | (1B) SII/III | (1C) Amb. | (1D) [4.5]E | (2) YSOc/giant | (3) Galc | (4) AGNc | (5) PAH |
|---|---|---|---|---|---|---|---|---|---|
|  |  | Probable Members[a] |  |  |  | Non-Members[b] |  |  |  |
| Flame Nebula | 642 | 399 | 113 | 198 | 88 | 18 | 0 | 67 | 8 | 168 |
| W40[c] | 4240 | 994 | 281 | 470 | 243 | 110 | 1380 | 618 | 1062 | 186 |
| RCW 36 | 190 | 132 | 43 | 31 | 58 | 3 | 0 | 2 | 1 | 55 |
| NGC 2264 | 1330 | 641 | 163 | 353 | 125 | 55 | 0 | 488 | 189 | 12 |
| Rosette Nebula | 1135 | 824 | 304 | 276 | 244 | 19 | 8 | 297 | 4 | 2 |
| Lagoon Nebula | 1108 | 570 | 148 | 236 | 186 | 5 | 492 | 30 | 0 | 16 |
| NGC 2362 | 1065 | 71 | 18 | 47 | 6 | 0 | 0 | 805 | 188 | 1 |
| DR 21 | 1498 | 746 | 223 | 308 | 215 | 55 | 0 | 399 | 195 | 158 |
| RCW 38 | 717 | 640 | 173 | 187 | 280 | 16 | 66 | 3 | 1 | 7 |
| NGC 6334 | 1211 | 685 | 292 | 187 | 206 | 59 | 494 | 11 | 0 | 21 |
| NGC 6357 | 1062 | 546 | 221 | 169 | 156 | 23 | 458 | 21 | 0 | 37 |
| Eagle Nebula | 1215 | 744 | 315 | 206 | 223 | 20 | 442 | 20 | 0 | 9 |
| M17 | 1137 | 142 | 55 | 40 | 47 | 6 | 941 | 42 | 0 | 12 |
| W3 | 184 | 181 | 50 | 52 | 79 | 1 | 0 | 1 | 0 | 2 |
| W4 | 1314 | 460 | 59 | 300 | 101 | 2 | 0 | 350 | 391 | 113 |
| Trifid Nebula | 540 | 292 | 116 | 107 | 69 | 21 | 227 | 11 | 0 | 10 |
| NGC 3576 | 790 | 220 | 72 | 80 | 68 | 3 | 501 | 31 | 3 | 35 |
| NGC 1893 | 1341 | 399 | 84 | 221 | 94 | 2 | 0 | 618 | 269 | 55 |
| Total | 20719 | 8686 | 2730 | 3468 | 2488 | 418 | 5009 | 3814 | 2311 | 899 |

[a] All probable members (MEM_flg = 1, regardless of SED_flg, see Table 3), should be considered candidate YSOs, further subdivided by most probable evolutionary stage (see column 12 in Table 3): envelope-dominated (S0/I), disk-dominated (SII/III), or ambiguous (Amb.). Sources with molecular shocks producing elevated 4.5 $\mu$m emission ([4.5]E) are a further subset of YSOc, predominantly S0/I.

[b] Sources not classified as members (MEM_flg = 0) are subdivided into the following groups (see column 9 in Table 3): YSOs or highly-reddened field giants that are not distinguishable from YSOs (YSOc/giant), candidate starburst/PAH galaxies (Galc), candidate obscured AGN (AGNc), and PAH nebular knots.

[c] W40 probable members include objects in the Serpens South molecular cloud.

Table 6. MIRES Catalog Tallies II: MYStIX X-ray Fields

| MYStIX Field | (1) All | (1A) S0/I | (1B) SII/III | (1C) Amb. | (1D) [4.5]E | (2) YSOc/giant | (3) Galc | (4) AGNc | (5) PAH |
|---|---|---|---|---|---|---|---|---|---|
| | | Probable Members (MPCMs) | | | | Non-Members | | | |
| Flame Nebula | 277 | 179 | 60 | 67 | 52 | 3 | 0 | 3 | 0 | 95 |
| W40 | 515 | 302 | 77 | 128 | 97 | 5 | 0 | 57 | 15 | 141 |
| RCW 36 | 190 | 132 | 43 | 31 | 58 | 3 | 0 | 2 | 1 | 55 |
| NGC 2264 | 805 | 523 | 145 | 272 | 106 | 55 | 0 | 209 | 61 | 12 |
| Rosette Nebula | 735 | 586 | 212 | 201 | 173 | 12 | 0 | 146 | 2 | 1 |
| Lagoon Nebula | 425 | 374 | 87 | 163 | 124 | 4 | 41 | 0 | 0 | 10 |
| NGC 2362 | 411 | 38 | 10 | 23 | 5 | 0 | 0 | 297 | 76 | 0 |
| DR 21 | 850 | 484 | 161 | 165 | 158 | 49 | 0 | 194 | 68 | 104 |
| RCW 38 | 105 | 94 | 36 | 19 | 39 | 2 | 5 | 0 | 0 | 6 |
| NGC 6334 | 404 | 324 | 130 | 100 | 94 | 33 | 67 | 2 | 0 | 11 |
| NGC 6357 | 487 | 389 | 146 | 132 | 111 | 11 | 73 | 3 | 0 | 22 |
| Eagle Nebula | 802 | 674 | 284 | 186 | 204 | 19 | 118 | 4 | 0 | 6 |
| M17 | 186 | 72 | 27 | 19 | 26 | 2 | 103 | 4 | 0 | 7 |
| W3 | 173 | 170 | 47 | 46 | 77 | 1 | 0 | 1 | 0 | 2 |
| W4 | 394 | 143 | 18 | 95 | 30 | 0 | 0 | 98 | 99 | 54 |
| Trifid | 181 | 140 | 54 | 56 | 30 | 12 | 38 | 1 | 0 | 2 |
| NGC 3576 | 181 | 114 | 33 | 45 | 36 | 2 | 63 | 1 | 1 | 2 |
| NGC 1893 | 1006 | 365 | 76 | 198 | 91 | 2 | 0 | 405 | 215 | 21 |
| Total | 8127 | 5103 | 1646 | 1946 | 1511 | 215 | 508 | 1427 | 538 | 551 |

Note. — See notes to Table 5; this Table is a subset, spatially restricted to the MYStIX X-ray fields-of-view (XFOV = 1; column 15 in Table 3). "MPCMs" are included in the catalog of Broos et al. (2013), which includes the subset of MIRES probable complex members spatially restricted to the MYStIX X-ray fields.

photometry available from UKIRT (King et al. 2013), while others have only *2MASS* photometry. These competing photometric sensitivity limits create large variations in completeness as a function of bolometric luminosity (a proxy for stellar mass) both between different target MSFRs and *even across a given MIRES field*. MIRES should provide a near-complete sample of the YSO population to sub-solar masses for targets resembling the relatively nearby NGC 2264 complex, with its deep UKIRT and IRAC photometric catalogs. By contrast, MIRES samples primarily the intermediate-mass (2–8 M$_\odot$) YSO population but is substantially incomplete even at solar masses in more distant targets resembling the Trifid Nebula, with only the shallower GLIMPSE Catalog available, limited further by brighter nebulosity and confusion from the dense field star population (see also Povich & Whitney 2010; Povich et al. 2011). In addition, the following important types of IRE sources will generally be missing from MIRES, given our selection criteria:

1. *Dense clusters in bright, compact H II regions.* Any areas where the [8.0] mosaics in Figure 9 saturate to white show MIR nebular background emission so extreme as to preclude the detection of the large majority of point sources over the *entire* luminosity range. M17 is among the worst offenders (Povich et al. 2009). Confusion in dense clusters also precludes detection of point sources at the 2″ IRAC resolution.

2. *Massive YSOs with saturated MIR emission or resolved disks/envelopes.* The NIR and MIR point source catalogs used for MIRES do not include photometry for saturated sources or resolved sources.

3. *Young stars with transitional disks.* Subject to intense, recent observational and theoretical study because of their connection to giant planet formation, disks representing the transition from primordial (optically thick, Stage II) to debris disks have SEDs showing strong IRE at wavelengths longer than 8 $\mu$m (e.g. Currie et al. 2009). While such objects are undoubtedly present in the MIRES fields, they most likely would manifest as marginal IRE sources, lost among overwhelming contamination by non-IRE stars showing spurious [5.8] or [8.0] IRE (see Section 3.2). Transition disk candidates could possibly be identified among the SEDs of X-ray selected MPCMs (see Appendix A).

The literature for identifying and classifying IRE sources based on *Spitzer*/IRAC photometry now spans a decade. As MIRES represents the most recent, and arguably most complicated, such methodology, it is worth comparing our classification results to the early, more straightforward approaches. Allen et al. (2004) showed that YSOs separate most cleanly from field stars on the [3.6] − [4.5] versus [5.8] − [8.0] IRAC color-color diagram, and they identified a box-shaped "disk domain" containing the locus of disk-dominated IRE sources. They showed that protostars are found redward of the disk domain, primarily in the [3.6] − [4.5] color, while normally-reddened field stars are found blueward of the disk domain in the [5.8] − [8.0] color. In Figure 10 we plot on the [3.6] − [4.5] versus [5.8] − [8.0] color space the 5496 MIRES detected in all 4 IRAC bands with uncertainties ≤0.1 mag. The large majority of Stage II/III disk-dominated YSOs are indeed located within the Allen et al. (2004) disk domain. Stage 0/I protostars dominate sources with $[3.6] - [4.5] > 0.75$ mag, and many of the reddest of these are [4.5]E candidate outflow sources.

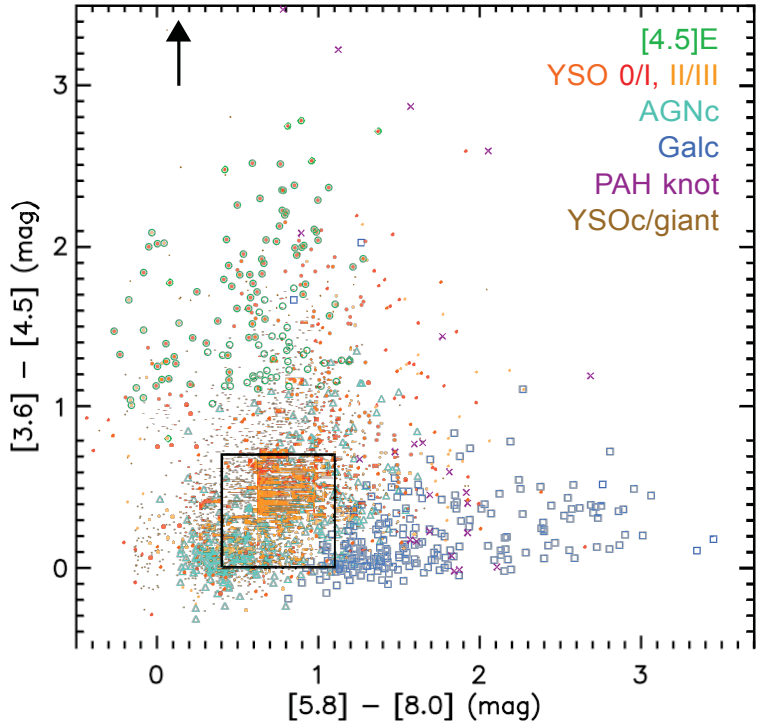

Fig. 10.— Color-color diagram used by Allen et al. (2004) to identify disk-bearing YSOs (within solid box) and protostars. The 5496 MIRES detected in all 4 IRAC bands with photometric uncertainties $\leq 0.1$ mag are plotted.

We do not, however, find any clean separation between Stage 0/I and Stage II/III sources in the color space of Figure 10 (or in any other color space, see Figs. 3 and 6), and the region of overlap is heavily populated by sources with ambiguous stage classifications. This is the largest difference between our SED-model-based classifications and the work of Allen et al. (2004), G09, and other authors using color-color diagrams (see also the discussion in Povich et al. 2011). Geometry and viewing angle often conspire to make protostars appear bluer than disk-dominated YSOs (1) in $[5.8] - [8.0]$ where the envelope produces a deep 9.7 $\mu$m silicate absorption feature, or (2) in $[3.6] - [4.5]$ where light from the central star scatters off of the envelope cavity walls (Whitney et al. 2003). The definition of Stage 0/I from RW06 includes both deeply-embedded protostars with red spectral indices at all IR wavelengths and more evolved objects retaining significant longer-wavelength ($\lambda \gtrsim 8$ $\mu$m) emission from infalling envelopes. Classifications based solely on 1–8 $\mu$m colors or SEDs cannot reliably distinguish the latter case from disk-dominated YSOs.

Figure 10 illustrates a clear disadvantage of relying solely on MIR colors for IRE source selection: YSOs do not separate cleanly from the various contaminating populations, especially obscured

AGN and reddened field giants. The work of the IRAC GTO team (including G09) and the Cores to Disks (c2d) survey team (notably Harvey et al. 2006) in cleaning extragalactic contaminants from IRE samples greatly ameliorates this fundamentally intractable problem. The MIRES targets include MSFRs at significantly greater distances and with far more variation in background reddening through the Galaxy compared to the regions commonly studied by these groups, requiring us to incorporate our SED-based dereddening into the G09 color cuts and to apply spatial clustering analysis to establish membership of faint MIRES in the more distant MSFRs.

Requiring detections in all 4 IRAC bands with high-precision photometry significantly reduces the number of IRE sources that can be identified and classified. Only 26.5% of all MIRES, including 40.2% of probable members, are plotted in Figure 10. To get around this limitation, G09 provide a "Phase 2" analysis using only $JHK_S$ plus [3.6] and [4.5] photometry to identify YSOs lacking detections at [5.8] and [8.0]. This analysis is more vulnerable to confusion between reddened field stars and YSOs than the G09 "Phase 1" analysis using all IRAC bands, and Povich et al. (2009) showed that even the Phase 1 color-color analysis would mis-classify large numbers of highly-reddened field stars (most likely giants that already have intrinsically redder IR colors) as YSOs in the inner-Galaxy M17 field. We find that an initial filtering based on fitting reddened stellar photospheres to SEDs, as employed in this work, its direct antecedents (Povich et al. 2009; Kang et al. 2009; Povich & Whitney 2010; Povich et al. 2011), or the parallel strategy employed by the c2d team (Harvey et al. 2006), is a critical step to mitigate otherwise overwhelming field-star contamination in dense, inner Galaxy fields.

With MIRES, we have analyzed a broad range of Galactic MSFRs in terms of distance, Galactic location, and depth of photometric data. MIRES provides both a unified strategy for identifying young stellar members of diverse MSFRs based on IRE emission and the basic lesson that there is no one ideal strategy. The central compromise for MIRES is the aggressive rejection of likely contaminants in favor of a more reliable sample of stellar members. Broos et al. (2013) combine MIRES members with the X-ray-selected MPCM samples, providing the basis for MYStIX follow-up studies, and future research should also return to the larger MIRES sample to find additional YSO members of the MYStIX MSFRs. Other follow-up studies could incorporate longer-wavelength survey data from *Spitzer*/MIPS or *Herschel* or use MIRES as a finding chart for observations with the Atacama Large Millimeter Array or the KMOS multiobject spectrograph on the Very Large Telescope. The MIRES catalog should provide a foundation, a starting point for follow-up studies of diverse phenomena related to massive star cluster formation, including protostellar outflows, circumstellar disks, and star formation triggered by massive star feedback.

S. R. Majewski and R. Indebetouw generously provided data from the *Spitzer* Vela–Carina Survey for this work. We thank M. R. Meade and B. L. Babler for providing point-source photometry for W3 via the GLIMPSE pipeline. We thank the referee, M. Gagné, for constructive comments that helped improve this work. M.S.P. was supported by an NSF Astronomy and Astrophysics Postdoctoral Fellowship under award AST-0901646 during the main analysis phase of this project. The MIRES catalog is based on observations from the *Spitzer Space Telescope*, which is operated by the Jet Propulsion Laboratory (California Institute of Technology) under contract with NASA. This publication makes use of data products from the Two Micron All-Sky Survey, which is a joint project of the University of Massachusetts and the Infrared Processing and Analysis Center/California Institute of Technology, funded by NASA and the NSF. This work is based in part on data obtained as part of the United Kingdom Infrared Telescope (UKIRT) Infrared Deep Sky Survey and in part data obtained in UKIRT director's discretionary time. UKIRT is operated by the Joint Astronomy Centre on behalf of the Science and Technology Facilities Council of the U.K. The MYStIX project is supported at Penn State by NASA grant NNX09AC74G, NSF grant AST-0908038, and the Chandra ACIS Team contract SV4-74018 (G. Garmire & L. Townsley, Principal Investigators), issued by the Chandra X-ray Center, which is operated by the Smithsonian Astrophysical Observatory for and on behalf of NASA under contract NAS8-03060.

*Facilities:* CXO (ACIS), Spitzer (IRAC), CTIO:2MASS, UKIRT (WFCAM)

## A. SED Classification Applied to IR Counterparts of MYStIX X-ray Sources

Identifying probable members of the target MSFRs is a fundamental goal of MYStIX. Analysis of IR counterparts matched to X-ray sources has proven to be a critical component of classifying X-sources and evaluating their probability of membership in MSFRs (Broos et al. 2011, 2013). Naylor et al. (2013) matched the various MYStIX *Chandra* catalogs to the more complete *Spitzer* Archive source lists (Benjamin et al. 2003, K13), presenting an opportunity to find new IRE counterparts to X-ray sources that were omitted from the MIRES analysis, which was based on the highly-reliable Catalogs to reduce contamination in our "blind," IRE-only search.

In this appendix we present results from our MIRES SED classification methodology as applied to all MYStIX X-ray sources with sufficient IR counterpart photometry available to support our analysis. The product of this analysis is the SED Classification of IR Counterparts to MYStIX X-ray sources (SCIM-X) online table, described in Table A1.[2] The following points were implemented in producing SCIM-X:

- The combination of X-ray emission and IRE is a powerful indicator of youth. We therefore used the more-complete IRAC Archive lists in lieu of the highly-reliable Catalog lists used for MIRES.

- Cross-matching was done by Naylor et al. (2013) in an X-ray centric fashion. It is possible to get a different NIR match to a given MIR source that appears in MIRES. We have not attempted to quantify how often this occurs, but we note that erroneous NIR–MIR crossmatches most likely will result in no valid SED fits (column 5 of Table A2).

---

[2] The acronym is a nod to the idea that we have "skimmed the cream" of the X-ray sources with the most available IR counterpart data.

- X-ray selection allows us to be more strict in our stellar atmosphere fits: $\chi^2/N_{\rm data} \leq 1$ is required for good fits. The goodness-of-fit criteria for RW06 model fits to IRE SEDs are unchanged. This allows for the identification of weaker IRE sources, or more evolved circumstellar disks, compared to MIRES.

- SCIM-X incorporates non-IRE (SED_flg = −2) and marginal-IRE (SED_flg = −1) source classifications (column 10 of Table A1). The former are more robust indicators that a star has an evolved/dispersed disk than in the case of $JHK_s$ color analysis alone (see Fig. A1), and the latter are far more likely to represent transitional disks or optically thin disks than the vast numbers of sources rejected for consideration from MIRES.

We identified sources with marginal IR excess emission (see also Section 3.2), where the excess appears in only the single IRAC [5.8] or [8.0] band, using the procedure of Povich et al. (2011). We use [$\lambda$] to denote magnitudes in the various IRAC bands and $\delta([\lambda_i] - [\lambda_j])$ for the uncertainties on the colors computed from the (minimum 10%) uncertainties on Catalog flux densities. Sources for which

$$[3.6] - [4.5] < \delta([3.6] - [4.5]) + E([3.6] - [4.5]) \quad (A1)$$

and there is no detection at longer wavelengths are classified as marginal IRE. The color excess used for the de-reddening was calculated as

$$E([3.6] - [4.5]) = A_V \frac{(\kappa_{3.6} - \kappa_{4.5})}{\kappa_V} = 0.0135 A_V,$$

where the $\kappa_\lambda$ are opacities given by the extinction law (Indebetouw et al. 2005) and $A_V$ is the maximum interstellar reddening observed to field stars in each MIRES field (Table 1). Sources with both [5.8] and [8.0] photometry available that fail Equation A1 above may be classified as significant IRE (SED_flg = 0 in MIRES and SCIM-X) *only* if they satisfy *both* of the following conditions:

$$|[4.5] - [5.8]| > \delta([4.5] - [5.8])$$

$$[5.8] - [8.0] > \delta([5.8] - [8.0]),$$

otherwise they remain classified as marginal IRE, rejected from MIRES and flagged in SCIM-X. The 1–4.5 $\mu$m SEDs of SCIM-X marginal IRE sources were re-fit with reddened stellar atmospheres (SED_model_type = 1 in column 9 of Table A1), ignoring any available [5.8] or [8.0] photometry.

The fraction of marginal IRE sources rejects from MIRES that represent actual YSO candidates can be estimated from SCIM-X. SCIM-X includes $487/2511 = 0.19$ marginal-IRE objects for each YSOc (Table A2). MIRES contains 13,695 YSO candidates (sum of Columns 1+2 in Table 2), implying an *upper limit* of $0.19 \times 13,695 = 2602$ sources excluded from MIRES that were legitimate YSO candidates. This is an upper limit because marginal-IRE objects that are legitimate YSOs should be over-represented in an X-ray-selected sample compared to an IRE-selected sample, which includes a higher fraction of more embedded objects. Even this upper limit represents only 2% of the 101,814 marginal-IRE sources originally found Table 2), implying that >98% of objects excluded globally by the marginal IRE filter were *not* YSOs.

The results of the SCIM-X classifications are summarized in Table A2. The first two (unnumbered) columns give the total number of of MYStIX X-ray sources and the number with sufficient IR counterpart photometry for SED classification in each target MSFR; across all targets an average of 37% of X-ray sources have counterparts in SCIM-X. The last column in Table A2 gives tallies of SCIM-X significant IRE sources that have no counterpart in MIRES; these 1504 sources increase the tally of IRE sources in the MYStIX X-ray fields by 18%, particularly in regions with high MIR nebulosity or dense clusters with MIR sources suffer crowding (for example in NGC 6334, M17, and W3), as either high backgrounds or crowding can cause an MIR source to be excluded from the highly-reliable Catalog lists. In some targets with both deep *Chandra* and deep *Spitzer* data (in particular NGC 1893, which has the deepest *Chandra* integration among the 18 MYStIX targets analyzed for MIRES; Feigelson et al. 2013) the increase is driven by sources classified AGNc (column 4). Note that for source-by-source cross-indexing of SCIM-X to MIRES, the MIR_NAME must be used, as the MIRES coordinates are based on the MIR source positions while the SCIM-X coordinates are based on the X-ray source positions.

SCIM-X sources are plotted on the $J - H$ versus $H - K_S$ color-color diagram in the top panel of Figure A1. This color-color diagram enjoys great popularity because it is based on NIR photometry accessible from the ground. It is especially useful for MYStIX because the UKIRT data have higher resolution and are less compromised by nebular background emission compared to *Spitzer* (King et al. 2013), hence for many X-ray sources UKIRT photometry is the only available counterpart photometry (Naylor et al. 2013). This color space presents an extended, diagonal locus of normally-reddened stars (between the parallel reddening vectors in the top panel of Fig. A1), and $K_S$-excess sources are located to the lower-right of this region. By color-coding the NIR sources based on SCIM-X classification, we confirm the earlier conclusions of Whitney et al. (2003), RW06, and others that while $K_S$ excess is a reasonably robust indicator of the presence of circumstellar dust, the *lack* of $K_S$ excess emission provides no useful constraints on circumstellar dust disks. Marginal IRE sources are found throughout the locus of normally-reddened stars, with a few showing modest $K_S$ excess emission.

The bottom panel of Figure A1 presents SCIM-X sources plotted on the same IRAC color space shown in Figure 10 for MIRES, but here the color coding distinguishes IRE from non-IRE, AGNc/Galc, and sources with failed SED fits. The evolutionary stage of YSOs is omitted for clarity. No marginal IRE sources satisfied the photometric uncertainty ≤0.1 mag criterion at both [5.8] and [8.0] for inclusion in this plot. We find that the simple application of the Allen et al. (2004) color cuts does a reliable job of separating YSOs from stellar photospheres when X-ray selection is employed as a pre-filter against contamination, although there is some overlap between the IRE and non-IRE populations. As was the case for MIRES, only a minority of SCIM-X sources can be

Table A1.  SCIM-X Online Table Format

|      | Column Label | Description |
|------|--------------|-------------|
| (1)  | Xray_Name | MYStIX X-ray source name[a] |
| (2)  | RAdeg | Right ascension of X-ray source (J2000, degrees) |
| (3)  | Dedeg | Declination of X-ray source (J2000, degrees) |
| (4)  | MIR_Name | Source name in original IRAC Archive (GLIMPSE or Kuhn et al. 2013)[b] |
| (5)  | NIR_label | UKIRT or *2MASS* catalog source matched to X-ray source |
| (6)  | IRmag | Magnitudes in 7 IR bands used for SED fitting: $J$, $H$, $K_S$, [3.6], [4.5], [5.8], [8.0] |
| (7)  | IRmag_err | Uncertainties on the 7 IR magnitudes used for SED fitting,[c] *reset to floor values*[d] |
| (8)  | NIRphot_cat | Provenance of near-IR source matched to IRAC source: 0=*2MASS*, 1=UKIRT, $-1$=none |
| (9)  | SED_model_type | Type of SED model fit to source: 0=reddened stellar atmospheres, 1=RW06 YSO models |
| (10) | SED_flg | Source classification flag: $-2$=stellar photosphere, $-1$=marginal IRE, 0=likely YSO, 1=starburst galaxy, 2=AGN, $-99$=no acceptable SED fits |
| (11) | SED_chisq_norm | $\chi^2/N_{\rm data}$ of best-fit SED model, number of data points fit is $3 \leq N_{\rm data} \leq 7$ |
| (12) | SED_AV | Visual extinction $A_V$ determined from $\chi^2$–weighted mean of all acceptable SED fits |
| (13) | SED_stage | Evolutionary Stage classification, RW06 YSO models: 1=Stage 0/I, 2=Stage II/III, $-1$=ambiguous, $-99$=unclassified[e] |
| (14) | MYStIX_SFR | Name of MYStIX target MSFR |

[a]Xray_Name should be used for cross-indexing SCIM-X with the X-ray classification and MPCM tables in Broos et al. (2013).

[b]MIR_Name should be used for cross-indexing SCIM-X with MIRES (Table 3), as there are many sources in common.

[c]Value of $-99.99$ means that flux was used as an upper limit for SED fitting.

[d]As described in Section 3.1, minimum uncertainty used for SED fitting was set in flux density space using $\delta F_i \geq 0.05 F_i$ for $JHK_S$, [3.6], and [4.5] and $\delta F_i \geq 0.10 F_i$ for [5.8] and [8.0]. For *original* photometric error bars we refer the reader to the appropriate original source catalogs (King et al. 2013, K13 or GLIMPSE).

[e]Sources with SED_flg $< 0$ were not fit with RW06 models and hence cannot be classified according to YSO evolutionary stage.

Table A2. SED-Based Classification Tallies for MYStIX X-ray Sources

|  | All X-ray | In SCIM-X | (1) Non-IRE | (2) Marg-IRE | (3) YSOc | (3A) S0/I | (3B) SII/III | (3C) Amb. | (3D) [4.5]E | (4) AGNc[a] | (5) Failed[b] | New IRE[c] |
|---|---|---|---|---|---|---|---|---|---|---|---|---|
| Flame Nebula | 547 | 225 | 72 | 13 | 131 | 33 | 50 | 48 | 3 | 0 | 9 | 21 |
| W40 | 225 | 163 | 60 | 15 | 80 | 20 | 31 | 29 | 0 | 1 | 7 | 10 |
| RCW 36 | 502 | 132 | 31 | 5 | 88 | 22 | 20 | 46 | 1 | 2 | 6 | 15 |
| NGC 2264 | 1328 | 724 | 340 | 61 | 281 | 53 | 160 | 68 | 10 | 14 | 28 | 47 |
| Rosette Nebula | 1962 | 1139 | 736 | 31 | 238 | 56 | 116 | 66 | 1 | 101 | 33 | 137 |
| Lagoon Nebula | 2427 | 982 | 602 | 56 | 253 | 26 | 172 | 55 | 2 | 2 | 69 | 96 |
| NGC 2362 | 690 | 425 | 254 | 20 | 29 | 0 | 26 | 3 | 0 | 113 | 9 | 123 |
| DR 21 | 765 | 321 | 135 | 32 | 122 | 37 | 39 | 46 | 6 | 18 | 14 | 42 |
| RCW 38 | 1019 | 204 | 149 | 7 | 39 | 7 | 10 | 22 | 0 | 0 | 9 | 19 |
| NGC 6334 | 1510 | 518 | 315 | 17 | 127 | 40 | 45 | 42 | 3 | 0 | 59 | 84 |
| NGC 6357 | 2360 | 1050 | 656 | 42 | 244 | 55 | 103 | 86 | 1 | 2 | 106 | 136 |
| Eagle Nebula | 2830 | 1176 | 780 | 51 | 239 | 58 | 100 | 81 | 3 | 6 | 100 | 49 |
| M17 | 2999 | 679 | 484 | 19 | 110 | 26 | 36 | 48 | 1 | 0 | 66 | 84 |
| W3 | 2094 | 487 | 285 | 22 | 164 | 31 | 66 | 67 | 1 | 6 | 10 | 100 |
| W4 | 647 | 305 | 153 | 37 | 66 | 5 | 42 | 19 | 0 | 46 | 3 | 58 |
| Trifid Nebula | 633 | 228 | 142 | 7 | 60 | 9 | 38 | 13 | 0 | 0 | 19 | 34 |
| NGC 3576 | 1522 | 328 | 224 | 24 | 66 | 10 | 33 | 23 | 0 | 9 | 5 | 36 |
| NGC 1893 | 1442 | 849 | 197 | 28 | 174 | 12 | 125 | 37 | 1 | 432 | 18 | 413 |
| Total (MYStIX) | 25502 | 9935 | 5615 | 487 | 2511 | 500 | 1212 | 799 | 33 | 752 | 570 | 1504 |

Note. — Across all of MYStIX (bottom row), 9365 (37%) of X-ray sources have IR counterparts with successful SED classifications, while 570 (2%) had SED fitting attempted but failed.

[a] Because our methodology does not robustly distinguish between the extragalactic classifications Galc and AGNc (see column 10 of Table A1), we here assume that an X-ray detection is strong evidence in favor of AGNc.

[b] Sources with sufficient photometric information to attempt SED fitting, but all model fits failed, as determined from the values in columns 9 and 11 of Table A1: SED_chisq_norm >1 or >4 for SED_model_type = 0 or 1, respectively. The most likely causes for an SED fit failure are strong IR variability or cases where the NIR and MIR matches to the X-ray source were not the same star.

[c] This column gives the numbers of new, significant IRE sources identified in each region by SCIM-X that are *missing* from MIRES.

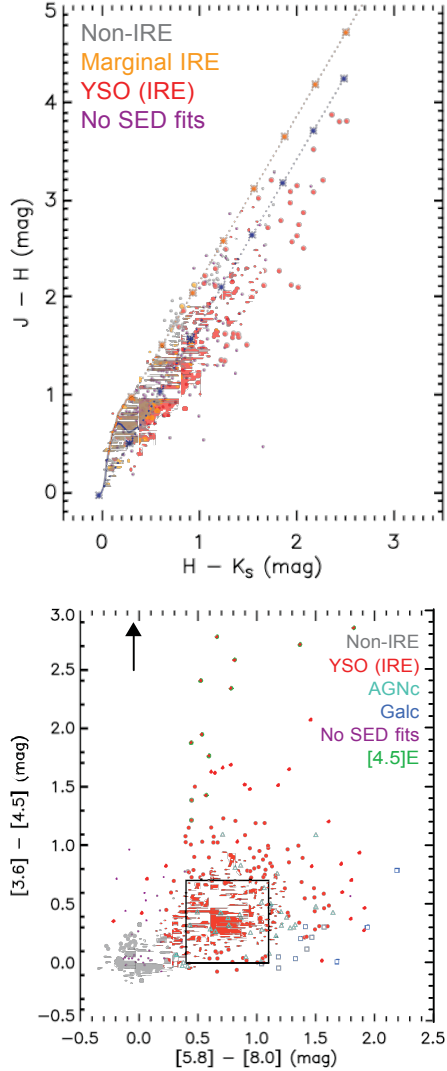

Fig. A1.— *Top:* $JHK_s$ color-color diagram for the X-ray-selected MYStIX source sample, including 1802 non-IRE, 157 marginal-IRE, and 632 significant-IRE sources detected in all 3 near-IR filters with photometric uncertainties $\leq 0.05$ mag and successful SED-based classifications. An additional 183 sources for which either no SED fitting was performed or no good SED fits were achieved are also plotted. 191 candidate starburst/PAH galaxies and 21 candidate AGN that satisfied the photometric criteria have been omitted for clarity; their colors strongly overlap the main locus of sources displayed. The loci of unreddened dwarfs and giants are plotted as blue and orange curves, respectively, and reddening vectors marked at $A_V = 5$ mag intervals extend from the end of each locus. *Bottom:* Mid-IR color-color diagram for the X-ray-selected sample, including all sources detected in all 4 IRAC bands with photometric uncertainties $\leq 0.1$ mag: 583 non-IRE, 990 significant-IRE (12 with 4.5E emission), 37 candidate AGN, 12 candidate starburst/PAH galaxies, and 38 sources for which no good SED fits were achieved.

classified using this color space.

## B. Descriptions of MIRES Populations Associated with Each MSFR

Here we provide brief, mostly qualitative descriptions of the MIRES populations associated with each target MSFR (distances quoted from Feigelson et al. 2013), as shown in Figure 9:

**The Flame Nebula** (Fig. 9*a*). The nearest MYStIX target ($d = 0.414$ kpc) presents a sightline toward the outer Galaxy, far from the Galactic plane. Contamination to MIRES from stellar and extragalactic sources is minimal, but very bright nebular emission produces numerous detections of PAH nebular knots, representing just over a quarter of all MIRES in this field (Table 5). The PAH nebular knots clearly trace the oval shape of the Flame Nebula, and the probable members, while commingled with the PAH knots, produce an elongated cluster with axis perpendicular to the long axis of the nebula. No cluster enhancement of MIRES coincident with the famous Horsehead Nebula (dark patch in the nebulosity south of the MYStIX field) is evident, but NGC 2023, the compact, bright nebula to the south, hosts its own clusters of MIRES members and PAH contaminants.

**W40 and Serpens South** (Fig. 9*b*). W40 presents a sightline toward the inner Galaxy, and with a deep *Spitzer* exposure (K13) this field is strongly contaminated both by non-associated stellar sources and extragalactic sources. Fortunately, the very large MIRES field provides an ample control field for establishing the baseline density of spatially distributed contaminants. Contamination from PAH nebular knots is also very strong in the central regions of the W40 MYStIX field. An additional complication is the presence of the Serpens South molecular cloud to the west. While Serpens South may be a foreground object located at only half of the 0.5 kpc distance to W40 (see Kuhn et al. 2010, and references therein), the superposition of the young stellar populations is sufficiently strong that we cannot draw a clean line spatially separating MIRES associated with Serpens South from MIRES members in W40. The MYStIX X-ray field, studied previously by Kuhn et al. (2010), is dominated by W40 members, but it also excludes a significant fraction of W40 MIRES members in the outer regions of the nebula.

**RCW 36** (Fig. 9*c*). The smallest (by far) MIRES field, RCW 36 ($d = 0.7$ kpc) is dominated by a dense, embedded cluster and associated bright nebulosity. RCW 36 presents the highest fraction of MIRES flagged as PAH knots (28.9%; Table 5).

**NGC 2264** (Fig. 9*d*). As one of the prototype MYStIX targets, NGC 2264 ($d = 0.913$ kpc) has been discussed previously in this work and by Feigelson et al. (2013). Extragalactic sources dominate the contaminants in MIRES. The spatial distributions of MIRES probable members agree qualitatively with those presented by Sung et al. (2009), who performed an independent photometric analysis of the same *Spitzer*/IRAC data to identify YSOs in NGC 2264. Two dense subclusters, with high fractions of Stage 0/I YSOs, are associated with the Spokes Cluster and the famous, optically-visible Cone nebula, although we note that the Cone itself (the southernmost in the chain of compact, bright IR nebulae) does not host a significant subcluster of MIRES probable members. There is also a more distributed population of MIRES probable members, dominated by Stage II/III YSOs, which includes the looser, possibly more evolved cluster associated with S Mon (northern end of the IR nebulosity).

**The Rosette Nebula and Molecular Cloud** (Fig. 9*e*). Our large, irregularly-shaped MIRES field does not fully encompass the Rosette Nebula, which spans 2° on the sky, hence there is no suitable control field on which to base a spatial distribution analysis for membership. Fortunately, the outer Galaxy sightline to the Rosette Nebula produces minimal contamination from stellar sources. Among MIRES flagged as Galc, the spatial distribution does not appear to be strictly uniform, instead exhibiting a tendency to concentrate with the YSOc. While there is a possibility that some fraction of Galc are in reality faint members of the Rosette complex at $d = 1.33$ kpc, without a control field we cannot establish a baseline density for spatial distribution analysis, and so only sources flagged YSOc are flagged as MIRES probable members. The majority of these are found within the chain of 6 *Chandra* pointings constituting the MYStIX field, which extends southeast from NGC 2244, the ionizing cluster of the Rosette H II region, through the long, filamentary Rosette molecular cloud. The fraction of Stage 0/I sources and degree of clustering is higher in the molecular cloud compared to NGC 2244. Ybarra et al. (2013) have studied the sequential progression of star formation activity in the Rosette molecular cloud using the same *Spitzer* data, applying a variant of the G09 methodology for identifying and classifying YSOs.

**The Lagoon Nebula (M8)** (Fig. 9*f*). Relatively nearby ($d = 1.3$ kpc) and large on the sky, the Lagoon Nebula presents a sightline passing only a few degrees from the Galactic center. The MIRES field is strongly contaminated by field stars, likely including evolved giants in the Galactic bulge observed through the high foreground reddening of the Galactic plane. Probable members represent just over half of MIRES for Lagoon (Table 5), with a large, central cluster flanked by numerous, smaller subgroups along the 8 $\mu$m boundaries of the H II region bubble. Most of the outer subgroups fall outside the boundaries of the MYStIX field. The fraction of Stage 0/I YSOs in most outer subgroups appears to be higher than that of the central cluster. Similar to the Rosette complex, the MIRES population is suggestive of sequential star formation propagating outward through an elongated molecular cloud (oriented along an east-west axis).

**NGC 2362** (Fig. 9*g*). The most evolved among the MYStIX clusters, NGC 2362 ($d = 1.48$ kpc) has dispersed its natal gas cloud, as evidenced by the complete lack of diffuse emission at 8.0 $\mu$m. Star formation has almost certainly ceased, and this field, dominated by extragalactic sources, contains the fewest YSOc among our 18 MIRES targets. The field was too small and too sparsely populated by YSOc to allow spatial clustering analysis, so all YSOc are classified as probable members, with an overdensity evident toward the field center. We caution that MIRES classified as Stage 0/I YSOs in NGC 2362 should be regarded with skepticism, as they may be faint sources with poor constraints from SED modeling, or even extragalactic contaminants misclassified as YSOc.

**DR 21** (Fig. 9*h*). DR 21 is a very young, massive star-forming cloud in the midst of the larger Cygnus X MSFR. Extragalactic contamination is high in the deep K13 *Spitzer* Catalog. While MIRES probable members with Stage 0/I classifications are strongly clustered in the DR 21 cloud, there is also a substantial distributed population of (predominantly Stage II/III) YSOc throughout Cygnus X (Beerer et al. 2010).

**RCW 38** (Fig. 9*i*). RCW 38 ($d = 1.7$ kpc) is a compact H II region producing very bright nebular emission in all IRAC bands. For MIRES we analyzed the shallow *Spitzer* Vela–Carina survey data in conjunction with 2MASS, as no UKIRT data were available for this target. Consequently, the MIR point-source sensitivity is very low throughout most of the MYStIX field, which is centered on the nebula. RCW 38 does appear to be associated with a much larger MSFR, and 85% of MIRES probable members are found in the extended membership field, outside the MYStIX field. These form two large, presumably older groups, dominated by Stage II/III YSOs, flanking the H II region to the northeast and southwest, plus a filamentary clustering, resembling DR 21, extending out of the field to the southeast. It is not clear whether these groups or clusters in the MIRES extended membership field are actually associated with RCW 38, but we choose to include them as probable members because they are significantly clustered with respect to the control field. Portions of these satellite clusters were also identified by Winston et al. (2011), who analyzed a smaller field of view using a deeper, targeted IRAC observation of RCW 38. These deeper data also included high-dynamic range photometry of the bright central core of the nebula, hence Winston et al. (2011) identified many more sources in the central cluster than are contained in MIRES.

**NGC 6334** (Fig. 9*j*). NGC 6334 is an enormous, elongated, MSFR extending 2° across the sky at $d = 1.7$ kpc, parallel to the Galactic plane. The MIRES field is constrained by the edges of the single, wide UKIRT field (King et al. 2013), so our MIRES probable members do not sample the entire MSFR, as defined by the 8 $\mu$m emission. Fortunately there are good control fields available where the clustered MIRES component falls off perpendicular to the long axis of the NGC 6334 complex, as the inner Galaxy sightline produces heavy contamination from unassociated stellar sources (YSOc). The MYStIX fields target the central clusters ionizing the optically-visible Cat's Paw Nebula, and here both crowding and the bright 8 $\mu$m nebular emission compromises the MIR point-source sensitivity. We note that, in spite of the high nebulosity, contamination from PAH nebular knots in minimal, as the GLIMPSE pipeline effectively rejects marginally resolved, compact sources. MIRES probable complex members reveal intense star-forming activity, dominated by the more readily-detected Stage 0/I YSOs, in numerous IR dark clouds criss-crossing the outer regions of the MSFR.

**NGC 6357** (Fig. 9*k*). Like NGC 6334, to which it appears to be joined in a single, giant molecular cloud complex at $d = 1.7$ kpc spanning several degrees across the inner Galactic plane (Russeil et al. 2010), NGC 6357 is a MSFR consisting of multiple massive clusters. It appears to be more evolved than NGC 6334, as the clusters have blown several H II region bubbles into the natal molecular cloud, and consequently the 8 $\mu$m nebular emission is less extreme, and MIR point-source detection more efficient compared to the case of NGC 6334. MIRES probable members trace the three main young stellar clusters, including the most famous, Pismis 24, as well as several satellite clusterings. NGC 6357 may also host a more distributed young stellar population (Wang et al. 2007), but the high contamination from YSOc/giants produced by a sightline only 7° from the Galactic center prohibits the identification of non-clustered MIRES probable members.

**The Eagle Nebula (M16)** (Fig. 9*l*). Indebetouw et al. (2007) previously studied the YSO population of the Eagle Nebula using the RW07 SED fitter applied to the GLIMPSE data, and is therefore a predecessor to the MIRES analysis of this MSFR. Compared to this earlier study, the MIRES catalog omits *Spitzer*/MIPS 24 $\mu$m photometry (but this could easily be added and has minimal impact on IRE source identification), includes deeper NIR photometry (King et al. 2013), and minimizes sample contamination from unassociated sources. The distributed component of YSOs reported by Indebetouw et al. (2007) disappears from the MIRES probable complex membership, but otherwise we find many of the same spatial features in the young stellar population, notably the absence of any significant, embedded clusters associated with the famous "Pillars of Creation" near the center of the MIRES field.

**M17** (Fig. 9*m*). The *Spitzer* YSO population of M17 ($d = 2.0$ kpc) was studied previously by Povich et al. (2009), using an earlier iteration of the MIRES analysis procedure. The MIRES catalog includes fewer probable M17 members, due to our more conservative selection criteria: adoption of the highly-reliable GLIMPSE Point Source Catalog versus the more-complete Archive used by Povich et al. (2009) and more stringent cleaning of spatially distributed contaminants. Among the MIRES fields, M17 is perhaps the most contaminated by unassociated YSOc clusters, especially toward the western field boundary (near the Galactic midplane). Only 142, or 13.1%, of the YSOc in the M17 field are MIRES probable complex members (Table 5). The members concentrate in three elongated groupings tracing molecular filaments along the western and northern boundaries of the M17 H II region, plus a fourth clustering to the north where the large IR bubble M17 EB interacts with an adjacent molecular cloud (Povich et al. 2009). NGC 6618, the massive young cluster responsible for ionizing M17, is swamped by MIR nebular emission and completely undetected in MIRES, in spite of its very high reported $JHK_S$ excess fraction (Hoffmeister et al. 2008).

**W3** (Fig. 9*n*). Several *Spitzer* studies exist in the literature of W3, a well-known MSFR in the outer Galaxy ($d = 2.04$ kpc), beginning with Ruch et al. (2007), and MIRES used the same IRAC GTO data for this target. Essentially all (181/184) MIRES in the W3 field are probable members, with spatial distribution agreeing with that reported by Ruch et al. (2007), although MIRES includes a larger number of sources because we did not require detection in all 4 IRAC bands. The IRAC GTO data do not completely cover the MYStIX X-ray fields, however they do encompass the majority of the young stellar population revealed in X-rays (Feigelson & Townsley 2008). Contamination is very low in this field, but unfortunately bright nebular emission severely limits the MIR point-source sensitivity near the youngest, embedded clusters W3 Main and W3(OH).

**W4** (Fig. 9*o*). W3 and W4 (and also W5, not a MYStIX target) belong to the same famous, enormous MSFR, which spans several degrees across the Perseus spiral arm in the outer Galaxy at 2.0 kpc. W4 is physically much larger than W3 and appears to be more evolved. The MIRES field samples only the central part of W4, the deep IRAC Catalog (K13) is dominated by Galc and AGNc; PAH nebular knots dominate the brighter diffuse emission regions, which otherwise might be mistaken for star-forming clouds. MIRES probable members are predominantly Stage II/III and themselves widely distributed, supporting the idea that W4 is relatively evolved.

**The Trifid Nebula (M20)** (Fig. 9*p*). As the second MYStIX prototype region, the often-photographed but relatively poorly-studied Trifid Nebula has been discussed previously in this work and by Feigelson et al. (2013), who note that its distance estimate was recently revised significantly outward, to 2.7 kpc, placing it behind the more evolved Lagoon Nebula, located <2° away in projection. The distribution of MIRES probable complex members reveal a rich extended star-forming environment, as the famous optically-visible nebula is threaded by one long, filamentary IR dark cloud on its western boundary (Lefloch et al. 2008). While the Trifid Nebula itself contains a central cluster of predominantly Stage II/III YSOs, the IR dark cloud hosts several tight clusters of Stage 0/I YSOs. The Trifid Nebula was studied previously by Rho et al. (2006) using *Spitzer* photometry, whose IRE selection criteria were based on the $[3.6]-[5.8]$ versus $[8.0]-[24]$ color-color diagram. None of the young MIRES clusters are readily apparent in the spatial distribution of IRE sources from this earlier work. The requirement of a *Spitzer*/MIPS 24 $\mu$m detection restricted their sample to relatively bright MIR sources, a significant fraction of which appear to be luminous, dust-rich AGB stars (Trifid presents a sightline intersecting the Galactic bulge).

**NGC 3576** (Fig. 9*q*). Like M17, NGC 3576 ($d = 2.8$ kpc) is a bright, compact H II region located along a complicated sightline passing through multiple spiral arms. NGC 3603, one of the most spectacular starburst clusters in the Galaxy, falls within the MIRES field 0.5° to the east of NGC 3576, but it is more than twice as distant. NGC 3603 was excluded from the control field for obvious reasons, but only a modest number of MIRES are found near NGC 3603, thanks to its great distance, high nebulosity, and extreme source crowding. It is unclear whether other MIR nebular features to the north and west in the MIRES field are molecular clouds associated with NGC 3576. We choose to include them within the MIRES extended membership field because of circumstantial evidence that dust pillars and illuminated cloud edges appear to be oriented toward NGC 3576. The association of MIRES probable members with NGC 3576 is more secure within the MYStIX field, which contains two main groupings of MIRES, a chain of compact, predominantly Stage 0/I clusters associated with the bright H II region (the ionizing cluster itself is not detected, of course) and a loose Stage II/III cluster to the north (Townsley et al. 2011).

**NGC 1893** (Fig. 9$r$). The most distant MYStIX target at 3.6 kpc, NGC 1893 is incompletely covered by the MYStIX and MIRES fields of view. The MIRES catalog reveals two elongated, young clusters of YSOc, apparently left in the wake of two bright, eroding dust pillars. Contamination from both extragalactic sources, which are not easily distinguished from faint, lower-mass YSOs is a general challenge for distant regions like NGC 1893, and one which MIRES only partially overcomes through spatial clustering analysis (Sections 5.1 and 5). The *Spitzer* data on NGC 1893 were previously analyzed by ), who identified 249 YSO candidates; MIRES contains 399 probable members, 21% of which are faint sources with GalC/AGNc flags.